**Anion Doping Driven Non-Ferroelectric-to-Ferroelectric Phase Transition in Epitaxial Y:HfO$_2$**

Soumyajyoti Mondal [a], Binoy Krishna De [a], Asraful Haque [a] , Shubham Kumar Parate [a], Arup Basak [a], Naushad Ahemad [a], Pramod Kumar Yadav [a], Kaushal Tiwari [b], Bhagwati Prasad [b], Matthew K. Sharpe [c], Catia Costa [c], Satheesh Krishnamurthy [c], Pavan Nukala [a,b,*]

[a] *Centre for Nano Science and Engineering, Indian Institute of Science, Bengaluru, 560012, India*.

[b] *Department of Materials Engineering, Indian Institute of Science, Bengaluru, 560012, India.*

[c] *Surrey Ion Beam Centre, University of Surrey, Guildford, Surrey, GU2 7XH, United Kingdom.*

* *Corresponding author* Pavan Nukala, e-mail: pnukala@iisc.ac.in

**Abstract**

Oxygen vacancies are often essential for stabilizing the orthorhombic ferroelectric phase in HfO$_2$, with cationic doping widely employed to introduce such defects. In contrast, systematic studies on anionic doping to induce ferroelectricity remains largely in nascent stages. Here, using epitaxial Y:HfO$_2$ films grown on ITO-buffered YSZ substrates that initially crystallize predominantly in the monoclinic non-polar phase, we demonstrate that post-deposition rapid thermal annealing in N$_2$ atmosphere at 900 °C enables nitrogen incorporation without disrupting epitaxy. As the annealing duration increases from 10 s to 2 min, the monoclinic phase diminishes, accompanied by the emergence of robust ferroelectric hysteresis and a corresponding increase in the orthorhombic phase fraction. Combining independent spectroscopic and compositional analyses, we experimentally establish that nitrogen preferentially incorporates into pre-existing neutral oxygen-vacancy sites, converting them into charged oxygen vacancies that drive the transformation from the non-polar monoclinic phase to the ferroelectric orthorhombic phase. Our epitaxial model platform therefore reveals an anion-mediated defect-engineering pathway for controlling ferroelectricity in Y:HfO$_2$, establishing nitrogen incorporation not merely as a chemical dopant, but as a route to fundamentally reconfigure the defect thermodynamics governing phase stability in fluorite ferroelectrics.

**Graphical Abstract**

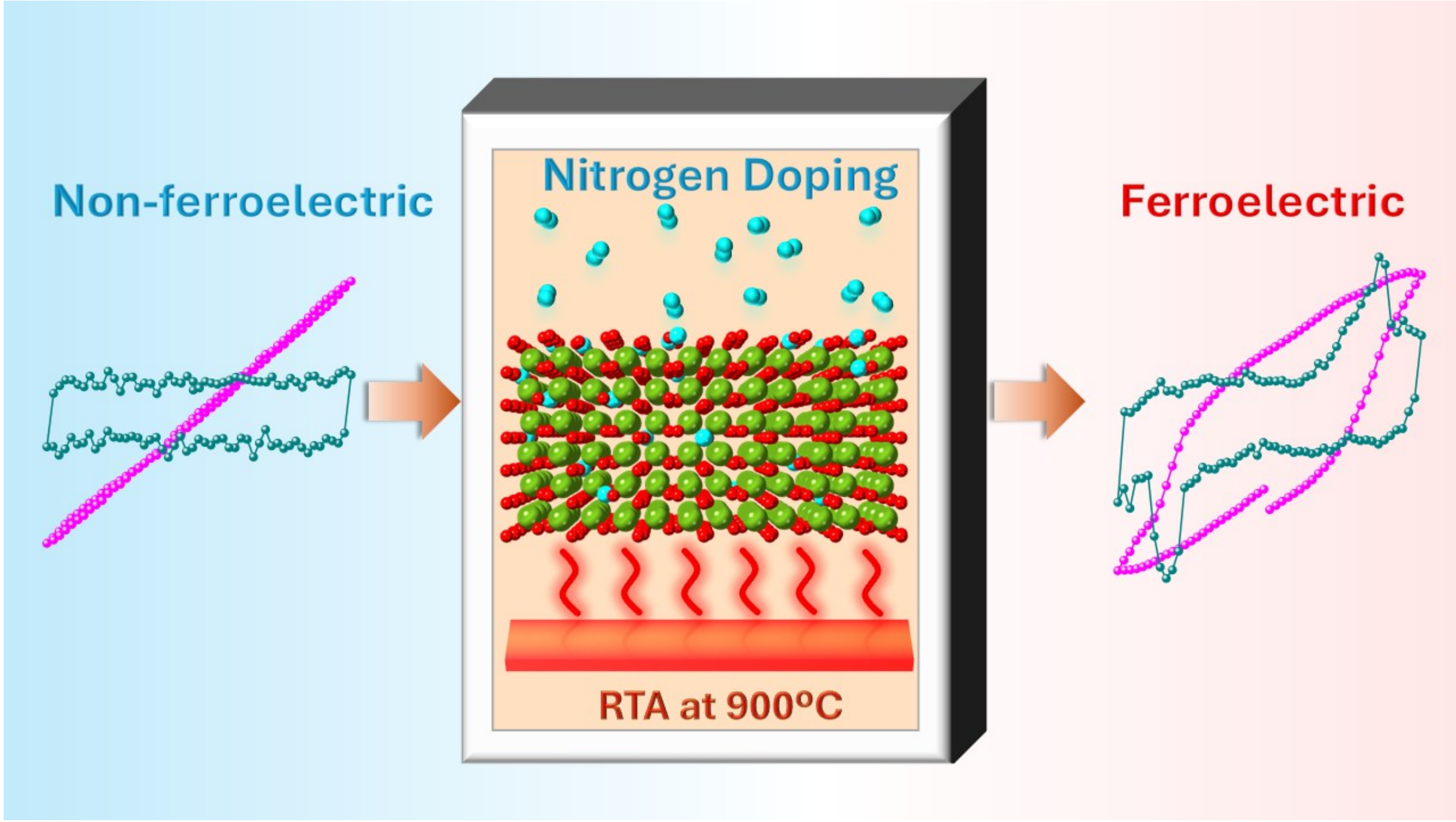

## 1. Introduction

Ferroelectric (FE) doped hafnium oxide ($HfO_2$) thin films have by now been established as a CMOS compatible solution for FE devices and associated technologies [1]. Contrary to conventional ferroelectrics, FE hafnia exhibits enhanced spontaneous polarization as film thickness decreases. This is due to the stabilization of metastable polar phases (orthorhombic $Pca2_1$ or rhombohedral R3 and R3m variants) at nanoscale over the bulk stable monoclinic $P2_1/c$ structure [2,3]. Precise control of this polymorph competition through cation doping [4–9], strain[10,11], interfaces [12,13], electrode materials [14–17], oxygen vacancies [18–23] , nanostructuring [24] and thermal processing [25] is essential for realizing robust ferroelectricity. Popular ways of synthesizing FE hafnia thin films include atomic layer deposition (ALD) [21,26,27], metal organic chemical vapor deposition (MOCVD) [28], RF magnetron sputtering [29], molecular beam epitaxy (MBE) [30,31], chemical solution processing (CSP) [32–35], and pulsed laser deposition (PLD) [36]. PLD is particularly advantageous for FE phase stabilization [37], offering precise stoichiometric transfer, epitaxial growth, and oxygen partial pressure control. These capabilities enable direct growth of highly crystalline films and stabilize the metastable orthorhombic and rhombohedral phases [10,38].

Ferroelectricity in hafnia is intrinsically coupled to oxygen vacancies and their dynamics. Across various techniques of sample growth, it has become increasingly evident that that the stabilization of polar and ferroelectric phases is often associated with an optimal concentration of oxygen vacancies [21,38,39]. A straightforward way of incorporating oxygen vacancies in $HfO_2$ is via heterovalent cation doping. A variety of heterovalent dopants larger than $Hf^{4+}$ such as Y, Sr, Ba, La have been shown to stabilize polar orthorhombic phase, while smaller cations like Mg or Al yield weaker responses, underscoring the importance of dopant chemistry [40,41]. Y and La can give ferroelectricity even large film thicknesses even up to 1 μm [42,43], and also in bulk [44].

Studies exploring the role of anion doping in ferroelectric hafnia remain relatively limited. In RF-sputtered $HfO_2$ thin films deposited in $N_2$-containing atmospheres on highly p-doped Ge substrates, an optimal level of nitrogen incorporation was reported to enhance ferroelectricity.[40,45]. It is speculated that oxygen vacancy formation due to the replacement of $O^{2-}$ with $N^{3-}$ ions is responsible for enhanced ferroelectric response [45]. Beyond these reports, however, rigorous mechanistic investigations establishing a direct correlation between anion incorporation, defect chemistry, phase evolution, and the resulting ferroelectric response remain largely absent.

In this work using epitaxial monoclinic rich, non-ferroelectric epitaxial Y:$HfO_2$ on ITO//YSZ, we systematically explore the effect of N doping on inducing ferroelectricity, through a simple high temperature RTA treatment under $N_2$ atmosphere. We hypothesize based on complementary x-ray Photoelectron Spectroscopy (XPS) and Time of Flight Elastic Recoil Detection Analysis (ToF-ERDA) experiments that N doping in monoclinic Y:$HfO_2$ transforms monoclinic phase into a polar orthorhombic phase by quenching neutral oxygen vacancies to create charged oxygen vacancies ($V_O^{\bullet\bullet}$), as per the following reaction:

$$3V_O^{\times} + N_2 = 2N_O' + V_O^{\bullet\bullet} \qquad \text{….. (equation 1)}$$

Our work fundamentally suggests that it is charged oxygen vacancies and their dynamics that is important for ferroelectric phase stabilization [46,47].

## 2. Results and Discussion

Y:$HfO_2$ epitaxial films were synthesized on ITO buffered YSZ (110) substrates, using Pulsed Laser Deposition (PLD) technique (see Methods). While substrate temperature, pressure, $O_2$ flow rates, and laser repetition rate were optimally fixed at 750 ºC, $1.2 \times 10^{-2}$ mbar, 5 sccm and 2 Hz respectively, varying the laser fluence from 2.7 to 3.6 J/cm$^2$ yielded films with different phases.

XRD θ-2θ scans at $\chi = 0°$ and $\chi \sim 35.26°$ are shown in **Figure 1a** and **1b** respectively for various laser fluences during Y:$HfO_2$ deposition. The scan for bare ITO is also shown for reference. Bragg peaks near 2θ~50° and 2θ~51.5° (**Figure 1a**) can be indexed as (220) planes of YSZ substrate and (440) planes of ITO bottom electrode respectively. The thin-film peak near 2θ~45° at the fluences of 3.1 and 3.6 J/cm$^2$ refers to the $(20\bar{2})$ planes of monoclinic Y:$HfO_2$. In case of sample grown at a lower fluence (2.7 J/cm$^2$, in red), the monoclinic$(20\bar{2})$ peak disappears and a faint shoulder peak near 2θ~50.5° appears, which corresponds to the orthorhombic {022} family of planes of Y:$HfO_2$ (shown by arrow). The evolution of the monoclinic phase (a higher volume phase) also increases the lattice parameter of the ITO layer due to counter clamping from the Y:$HfO_2$ layer. Conversely, ITO peak position also can indicate the major phase in the Y:$HfO_2$ layer (orthorhombic or monoclinic).

As the orthorhombic reflection is partially obscured by the substrate peak in on-axis scans, off-axis θ-2θ scans at $\chi$= 35.26° was employed as additional phase identification method. Bragg peaks near 2θ~30.1° and 2θ~30.6° can be indexed as (111) planes of YSZ substrate, and (222) planes of ITO bottom electrode respectively. For all the samples we observe a very broad (111) monoclinic peak between 2θ~28º and 29º. In case of sample grown at lower fluence (2.7 J/cm$^2$) the monoclinic peak is suppressed significantly. The (111) planes of the orthorhombic Y:$HfO_2$ also should appear at 2θ~30.1º, which is also masked by the intense substrate peak. However, the Laue fringes (shown by dashed rectangle) around the substrate peak (on the left) clearly indicate the presence of the masked orthorhombic phase peak, in addition to attesting for the high crystalline quality of the film. Notably, the sample grown at a lower fluence exhibits a narrower fringe width compared to the films deposited at higher fluences. This variation in the Laue fringe spacing indicates that a higher volume fraction of the secondary monoclinic phase effectively disrupts the structural continuity of the film, leading to a smaller coherent crystallite size for the ferroelectric orthorhombic phase. These results show that the ferroelectric orthorhombic phase is stabilized at lower fluences, whereas monoclinic-rich films are obtained at higher fluences. We speculate that at lower fluences, lighter species such as oxygen within the PLD plume reach the substrate in smaller quantities, which leads to the creation of oxygen vacancies (in addition to the ones created by Y doping) to stabilize the ferroelectric phase in hafnia-based ferroelectrics. HAADF-STEM measurements reveal that the thickness of Y:$HfO_2$ and ITO to be ~10 nm and ~30 nm respectively, and that the interface between ITO and Y:$HfO_2$ is very smooth and coherent (see Supplementary Information **Figure S1**).

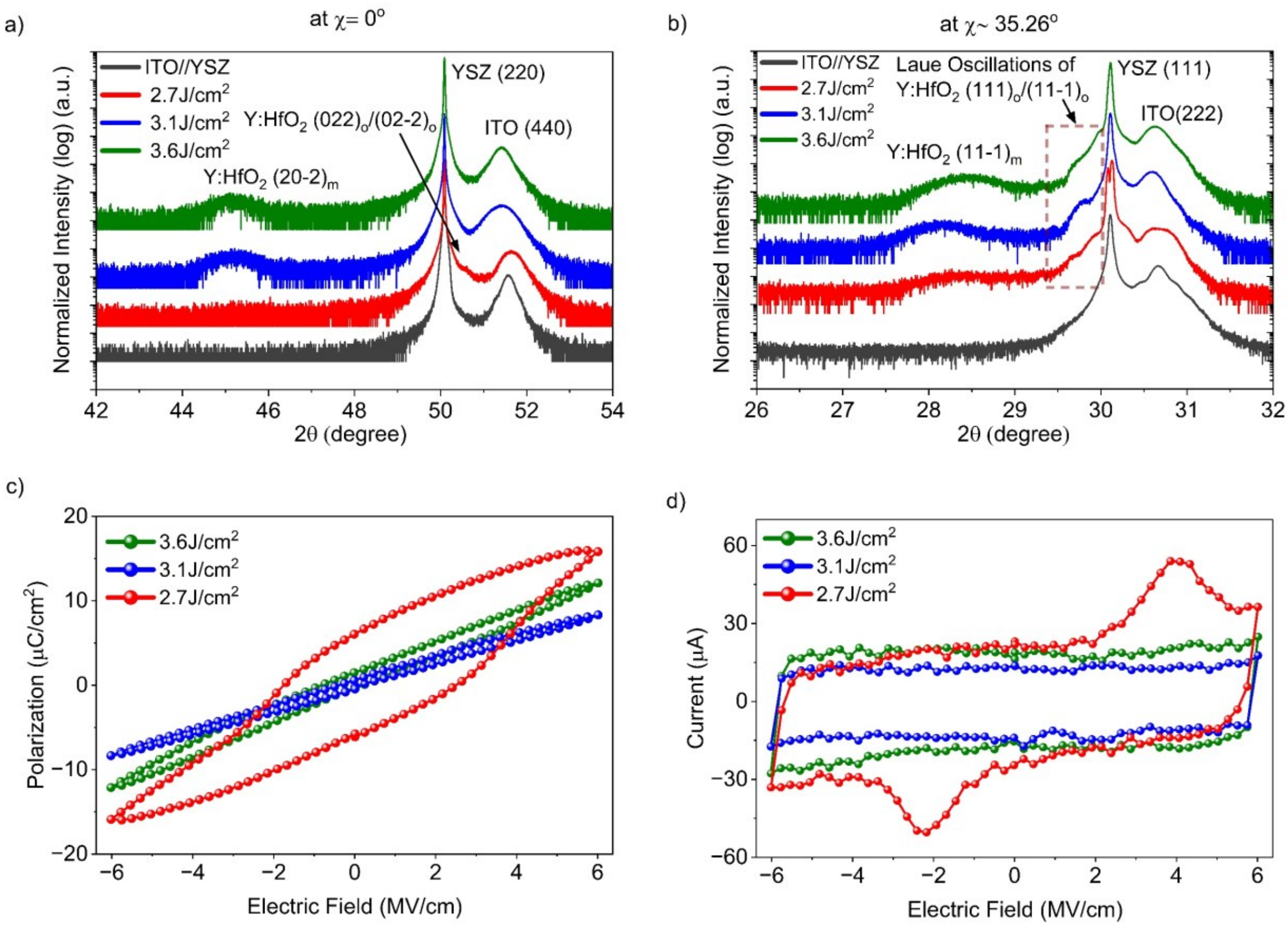

**Figure 1**. **Structural and electrical characterization of Y:$HfO_2$ films deposited at various laser fluences.** a) XRD θ-2θ scan at χ= 0º, b) XRD θ-2θ scan at χ= 35.26º, c) Polarization and d) Instantaneous current as function of electric field measured at 20 kHz frequency for the devices interfaced with W top electrodes.

To characterize the dielectric and ferroelectric properties of these films, we fabricated capacitors using TiN and W as the top electrodes, with their diameter varying from 30 μm to 50 μm. Polarization (P) and corresponding instantaneous current (I) as a function of electric field (E) for samples grown at various laser fluences and interfaced with W top electrodes are depicted in **Figures 1c** and **1d**, respectively. The devices on the films which are deposited at 3.6 and 3.1 J/cm$^2$ laser fluences exhibit pure dielectric behavior, however, film deposited at 2.7 J/cm$^2$ laser fluence exhibits ferroelectric behavior. In case of ferroelectric film with W top electrode, $2P_r$ is ~12 μC/cm$^2$ albeit with some imprint field of 0.9 MV/cm. However, with TiN as the top electrode, we measure a lesser value of $2P_r$ (9 μC/cm$^2$) and imprint field of 0.4 MV/cm (**Figure 2a** and **2b**). In both cases, there is no wake-up effect. A detailed device performance data for both electrodes with various frequencies and electric fields is given in Supplementary Information (**Figure S2** and **S3**).

To understand the differences between the capacitors with W and TiN as the top-electrodes, we performed HAADF-STEM imaging and EDS of FIB cross-sectioned samples (**Figure 2c**-**e**) on devices cycled for ~50 times, and compared them with control samples. For the capacitors with TiN as top electrodes, we see that entire TiN electrode is oxidized (see EDS spectra in **Figure 2c**), with more oxygen signal at the interface and lesser away from the interface, both larger than the control sample. This clearly suggests field-cycling induced oxygen migration, which can result in lesser imprint (**Figure 2b**). However, capacitors with W as the top electrode with a clean interface in the pristine state, clearly show an amorphous $WO_x$ (~1.5 nm) layer between the electrode and the Y:$HfO_2$ film after cycling (**Figure 2d** and **2e**). This layer can give rise to polarization pinning, trap charges and associated imprint effects.

Next, we performed endurance measurements at very low frequency of 20 kHz (and triangular pulse amplitude of 4.5 V) on capacitors with TiN and W top electrodes (Supplementary Information **Figure S4**). Note that frequencies such as 1 MHz are typically used for endurance measurements, and low frequency measurements show fatigue effects at lower number of cycles. From the data shown in Supplementary Information **Figure S4c**, we clearly see that the TiN electrode exhibits superior cyclability (>10$^7$ cycles even at 20 kHz), whereas devices with W top electrodes deteriorate with a smaller number of cycles, attributable blockage of oxygen migration (**Figure 2e**) by the top electrode. Our results are consistent with the observations on epitaxial $Hf_{0.5}Z_{0.5}O_2$ on LSMO//STO from reference [48], where an oxygen blocking electrode such as Au leads to an orthorhombic to monoclinic phase transformation with cycling attributed to oxygen accumulation in the hafnia layer. However, using an oxygen conductor such as LSMO as the top electrode doesn’t show this effect, resulting in superior cyclability [48].

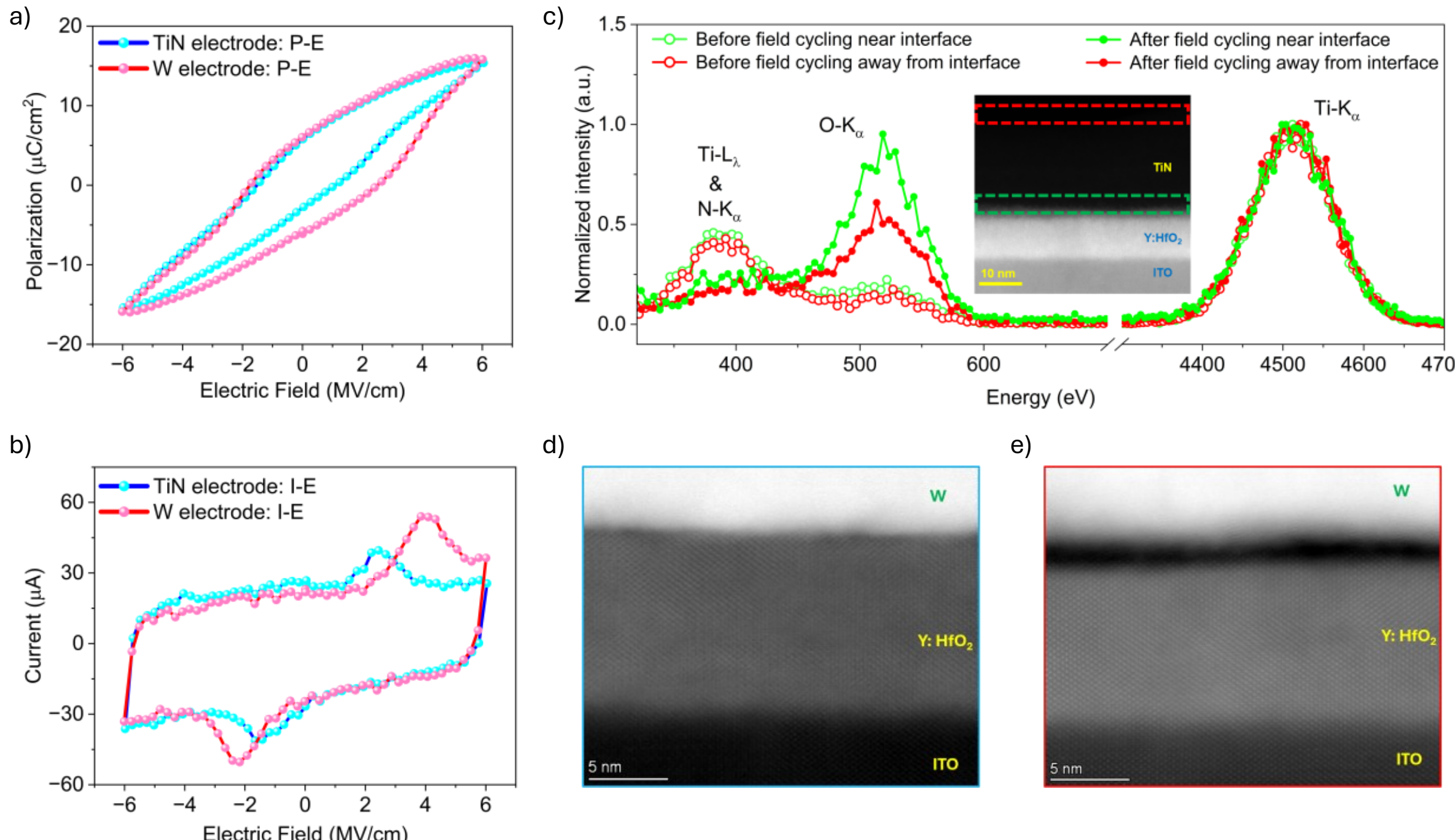

**Figure 2.** Effect of the top electrode on ferroelectric properties. (a) Polarization-electric field (P-E) and (b) current-electric field (I-E) loops for devices with TiN and W top electrodes. (c) TEM-EDS spectra taken from regions near and away from the top electrode/Y:$HfO_2$ interface, before and after field cycling. A representative HAADF-STEM image (inset) shows the positions where we collected the spectra. Cross-sectional HAADF-STEM images of the sample with a W top electrode (d) before and (e) after field cycling.

Rapid thermal annealing (RTA) under $N_2$ atmosphere is commonly used to obtain ferroelectric phases in doped $HfO_2$ [49,50]. For ALD grown samples, typically a top electrode such as W or TiN is deposited on the hafnia film, and RTA is performed on the capacitor structures [14,49,50]. The effect of the top electrode is better understood in terms of mechanical confinement during crystallization (the clamping effect), which restricts out-of-plane expansion and suppresses the transition to the larger-volume monoclinic phase in favor of the denser orthorhombic phase, [51], rather than a simple thermal expansion mismatch mechanism [52]. Even in PLD grown samples (Funakubo et al. [53–55]), RTA under $N_2$ atmosphere has been commonly used as a post-deposition protocol before top electrode deposition [53–55]. However, the role of RTA, particularly in $N_2$ ambience is not discussed in those works.

On our non-ferroelectric monoclinic samples, next we asked the question whether RTA at high temperature can N dope into hafnia, and whether N doping can induce non-FE to FE phase transformation. To this end, we systematically varied the time of exposure of our epitaxial monoclinic films to $N_2$ ambience in the RTA chamber from 10 sec to 4 min at 900 °C.

The XRD θ-2θ scan at $\chi$=0° (**Figure 3a**) shows that the monoclinic peak (at 2θ ~ 45°) reduces after 2 min of RTA under $N_2$ atmosphere. Similarly, XRD θ-2θ scan at $\chi$~35.26° (**Figure 3b**) shows the significant decrease of monoclinic phase (at 2θ ~ 28.5°) and emergence of orthorhombic peak (2θ ~ 30°- 30.1°) (**Figure 3b** inset). However, we observed reappearance of monoclinic phase after 4 min of RTA, consistent with readuction in polarization **(Figure S5)**.

We then performed P-E loops (and corresponding instantaneous current-voltage measurements) on all these samples by interfacing with W electrode subsequent to RTA treatment (**Figure 3c** and **3d**). We deliberately avoided TiN electrodes, despite their well-known advantage in improving cycling endurance, to isolate the effect of nitrogen introduced during the 900 °C RTA process. This choice eliminates ambiguity arising from possible nitrogen diffusion from the electrode during electrical cycling, ensuring that the observed ferroelectric response can be attributed to process-induced nitrogen incorporation rather than field-driven electrode-mediated effects.

While the dielectric hysteresis increases with RTA exposure upto 1 min, a clear ferroelectric loop emerges in samples exposed optimally for 2 minutes (**Figure 3c** and **3d**) and decreases for longer exposure (see Supplementary Information **Figure S5**). It is important to note that samples exposed to RTA at 900 ºC under Ar and $O_2$ atmosphere for 2 minutes did not show any reduction in the monoclinic phase (supplementary information, **Figure S6**), nor showed any ferroelectricity, illustrating that both the thermal treatment and the nitrogen atmosphere are important to induce ferroelectricity. The 2 min RTA-treated film demonstrates stable remanent polarization up to approximately $10^5$ cycles, after which fatigue behavior is observed (Supplementary Information **Figure S7**). While our films exhibit lower measured $P_r$ values compared to the theoretically predicted [56,57] spontaneous polarization for orthorhombic hafnia systems, the physical reasons for this discrepancy are detailed in Supplementary Information **Text S1**.

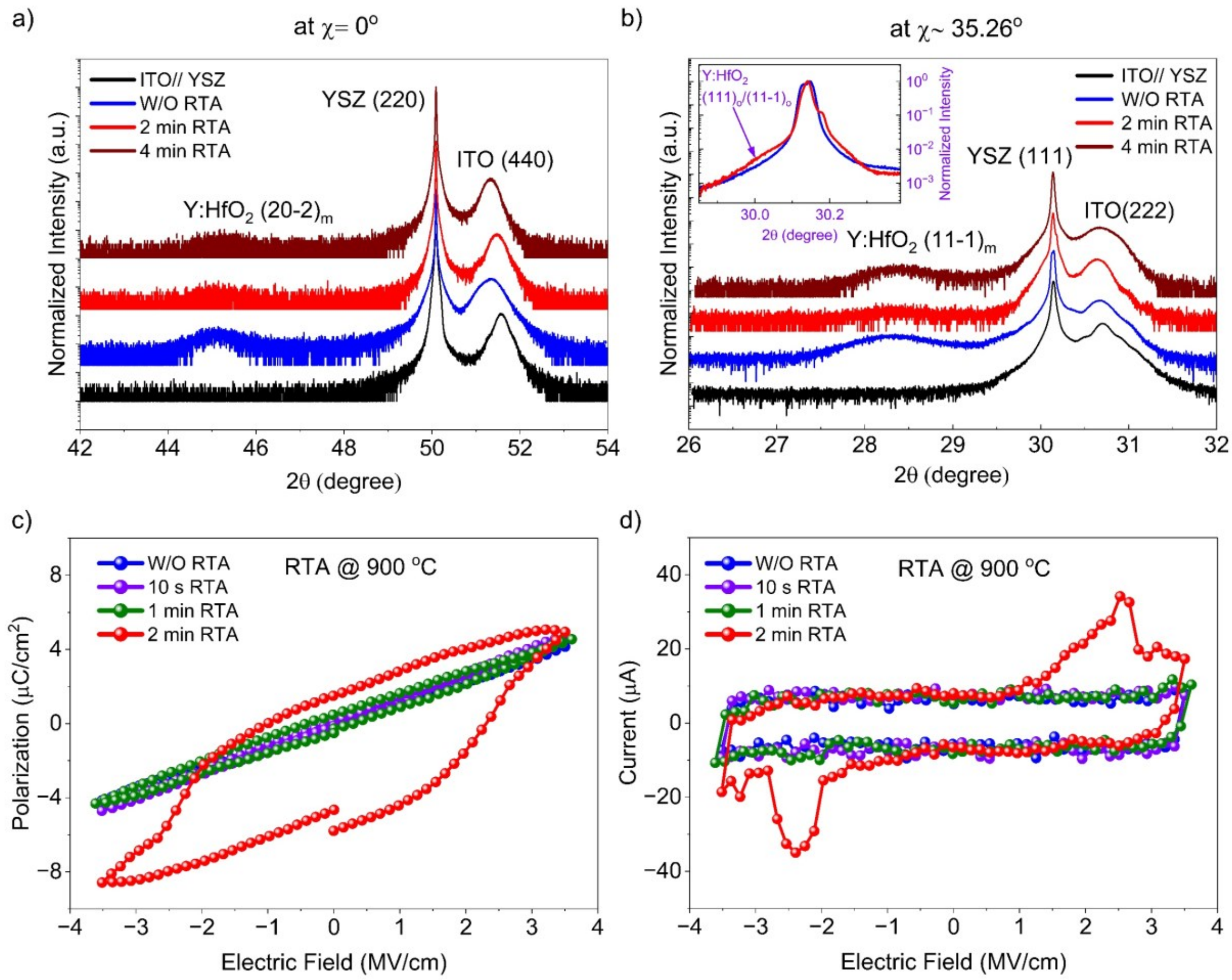

**Figure 3:** **Effect of rapid thermal annealing on as grown non-ferroelectric sample.** XRD θ-2θ scan for the samples annealed under $N_2$ ambience for various exposure duration a) at $\chi$=0°, b) at $\chi$=35.26°. The inset in (b) highlights the emergence of the orthorhombic peak near 2θ=30° for the sample annealed for 2 minutes compared to the unannealed sample.c) Polarization and d) Instantaneous current as a function of electric field for various durations of RTA exposure.

To understand the N incorporation as a function of RTA exposure, we performed XPS analysis concentrating on the N 1s peak (bonded with Hf) at binding energy of 397-398 eV (**Figure 4 a**) [58,59]. The Y 3s peak (at ~ 394-395eV) appears very close to N 1s peak [60]. Hence to deconvolve the N peak, we fit the Y 3d peak ~158 eV and calculated the intensity of the Y 3s. We deconvolve the spectra around 390-402 eV into two peaks, N 1s and Y 3s, where the area of the Y 3s peak is fixed to the calculated value as listed in **Table S2**. A detailed procedure for peak fitting is given in the Supplementary Information **Text S2** and **Figure S8**. The N to total cation ratio obtained thus (Supplementary Information **Figure S9 (inset)**) increases with increasing RTA exposure from 10 sec to 1 min, and saturates subsequently (within an error, see **Text S3**), indeed showing that RTA is a powerful tool to induce N doping in oxide thin films. We also note that O to total cation ratio also increases with RTA (**Figure 4c**). These observations are also consistent with time-of-flight elastic recoil detection analysis (**Figure S10** and **Table S5**), an independent composition verification method (**Figure 4d** and **e**). An increase in O concentration is not consistent with the mechanism proposed in an earlier work [45] on N doping, where $N^{3-}$ substitutes $O^{2-}$ sites, leading to the creation of oxygen vacancies ($V_O^{\bullet\bullet}$), overall reducing the O content.

In addition, our XPS results also show a clear blue shift in the binding energies of Hf 4f (**Figure 4g**), and O 1s lattice peak ($O_{P2}$ -1s), and an increase in the intensity of $O_{P1}$ -1s (**Figure 4f**), which is related to the presence of charged oxygen vacancies ($V_O^{\bullet\bullet}$). All these observations are consistent with the hypothesis that N gets incorporated into neutral oxygen vacancy sites ($V_O^{\times}$), generating acceptor defect, and a charged oxygen vacancy ($V_O^{\bullet\bullet}$), as illustrated in equation 1.

$$3V_O^{\times} + N_2 = 2N_O' + V_O^{\bullet\bullet} \qquad \text{..... (equation 1)}$$

The monoclinic phase of Y: $HfO_2$ consists of neutral oxygen vacancies ($V_O^{\times}$), which are complexes of $2Hf_{Hf}' - V_O^{\bullet\bullet}$. Hf is in $3^+$ oxidation state in this complex. Two neutral vacancies are filled by N, giving rise to two single charged $N_O'$ acceptors and one fully charged oxygen vacancy ($V_O^{\bullet\bullet}$), decreasing the total number of oxygen vacancies. All-in-all, N content increases, concomitantly with oxygen content. Furthermore, the presence of charged vacancies ($Hf^{3+}$ is oxidized to $Hf^{4+}$), shifts the binding energy of both $O_{P2}$ -1s and Hf 4f peaks to higher values (**Figure 4f** and **g**), in addition to increasing the intensity of $O_{P1}$ -1s peak corresponding to $V_O^{\bullet\bullet}$. These $V_O^{\bullet\bullet}$ (coupled with the acceptor $N_O'$) transforms monoclinic phase into orthorhombic phase. So far, there have been theoretical predictions about the effect of charged vacancies on stabilization of polar phases [46,47]. This work shows a clear experimental proof in favor of these predictions [46,47].

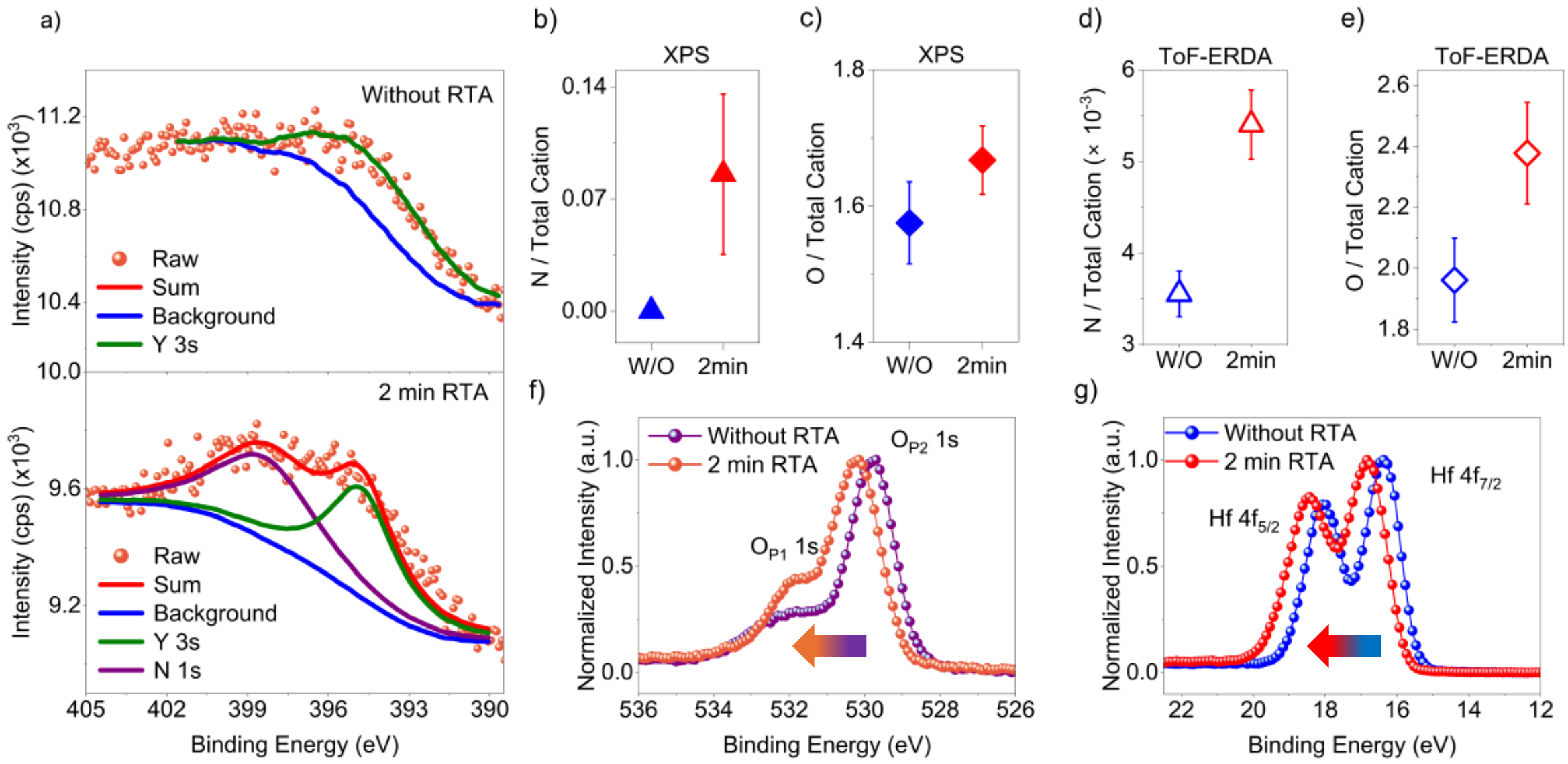

**Figure 4:** Chemical composition and bonding state analysis of the as-grown and 2-minute rapid thermal annealed (RTA) films. (a) X-ray photoelectron spectroscopy (XPS) spectra of the N 1s and Y 3s core levels for the film without RTA (top) and after a 2-min RTA treatment (bottom). Deconvolution reveals an N 1s peak in the RTA-treated sample. N/total cation and O/total cation ratios obtained from (b, c) XPS and (d, e) ToF-ERDA. Normalized XPS spectra of the (f) O 1s and (g) Hf 4f core levels before and after RTA. The arrows denote a shift toward higher binding energies for the 2-min RTA sample.

Finally, although the dissociation of molecular $N_2$ requires a large energy (~10 eV) [61], nitrogen incorporation during RTA is nevertheless feasible because it occurs preferentially at pre-existing oxygen-vacancy defects rather than through direct dissociation in the perfect lattice. First-principles studies have shown that oxygen vacancies in $HfO_2$ provide energetically favourable trapping sites for nitrogen [61]. In epitaxial Y:$HfO_2$, the high concentration of oxygen vacancies introduced by Y doping, together with kinetically trapped defects generated during PLD growth, likely facilitates nitrogen uptake during annealing. This interpretation is fully consistent with our compositional, spectroscopic, and structural observations, which collectively indicate that nitrogen incorporation modifies the oxygen-vacancy population and stabilizes the ferroelectric orthorhombic phase.

## Conclusion

In summary, we demonstrated a direct, non-ferroelectric to ferroelectric phase transition in epitaxial Y-doped $HfO_2$ thin films grown on ITO//YSZ(110) substrates. We achieved this transformation through a simple post-deposition rapid thermal annealing (RTA) process under a nitrogen atmosphere. Comparative control experiments under inert Argon and oxidizing oxygen environments explicitly confirm that nitrogen incorporation, rather than the thermal budget alone, acts as the primary driver for this structural phase transition.

Our comprehensive structural and compositional analyses utilizing off-axis XRD, XPS, and ToF-ERDA reveal a unique and fundamentally distinct phase stabilization mechanism. During the RTA process, nitrogen incorporates into the hafnia lattice and preferentially occupies pre-existing neutral oxygen vacancy $V_O^{\times}$ sites. This specific reaction quenches neutral vacancies and simultaneously generates charged oxygen vacancies ($V_O^{\bullet\bullet}$). We provide the first direct experimental proof that these charged defect complexes significantly suppress the thermodynamically stable monoclinic phase and stabilize the polar orthorhombic phase, enabling robust ferroelectric switching.

This defect engineering process preserves the single-crystalline epitaxial registry of the film. Consequently, this work establishes a model system for understanding structure-property correlations in fluorite-structure ferroelectrics and exploring defect engineering in the broader field of oxide electronics.

## 3. Methods

***Substrate and bottom electrode selection:*** (110) YSZ substrates were selected for the growth of Y: $HfO_2$. We used epitaxial ITO buffer layer as the bottom electrode between YSZ and the hafnia film. One unit cell of ITO lattice matches well with the four-unit cells of both YSZ and Y:$HfO_2$ maintaining lattice matching and minimizing interfacial defects (see in Supplementary Information, **Figure S11**).

***Y:HfO₂ and ITO target preparation:*** Commercially available ITO ($In_2O_3$:$SnO_2$ ~ 90:10) target was ablated for bottom electrode deposition. A target of ~7% Y dopped $HfO_2$ was prepared by solid state synthesis by mixing 99.99% $Y_2O_3$ and 99.95% $HfO_2$ powders. XRD θ-2θ scan of the powder scratched from the Y:$HfO_2$ and ITO PLD targets are shown in Supplementary Information **Figure S12**

***Synthesis of Y:HfO₂/ITO films:*** Base pressure of the PLD chamber was $1 \times 10^{-7}$ mbar. ITO films were deposited at 700 ºC substrate temperature, $1.2 \times 10^{-3}$ mbar chamber pressure, 5 sccm $O_2$ flow, with 2.1 J/cm$^2$ laser fluence and 3 Hz repetition rate. Y:$HfO_2$ films were deposited at 750 ºC substrate temperature, $1.2 \times 10^{-2}$ mbar chamber pressure, 5 sccm $O_2$ flow, with a range of laser fluence from 2.7 to 3.6 J/cm$^2$ and 2 Hz repetition rate. Target to substrate distance was maintained at 4.5 cm and substrate rotation was performed during film deposition for both films to maintain the thickness uniformity. After the deposition, films were held at 600 ºC for 30 min at $1.5 \times 10^{+2}$ mbar $O_2$ pressure, then cooled down to room temperature.

***RTA process:*** We utilized a custom-built RTA chamber (**Figure S13**) equipped with a thermocouple, vacuum pump, pressure gauge, mass flow controller, and chiller. Prior to annealing, the chamber was evacuated to a base pressure of $2\times10^{-3}$ mbar. An $N_2$ gas flow was then introduced at 200 sccm, maintaining a total pressure of $5\times10^{+2}$ mbar during the process. The RTA was performed at 900 ºC for various durations. Both the heating ramp rate and the initial cooling rate (down to ~200 ºC) were 50 ºC/s, after which the system cooled naturally to room temperature at a slower rate.

***Electrode lithography and sputtering:*** The top electrodes were defined using photolithography. The samples were solvent-cleaned and spin-coated with AZ5214E photoresist. Device patterns were defined using a Heidelberg µPG 501 direct-write photolithography system and developed in AZ 726 MIF developer to form circular pads with diameters of 30 µm and 50 µm. Following lithography, the W and TiN top electrodes (100 nm thickness) were deposited by DC sputtering and DC reactive sputtering, respectively. The sputtering system achieved a base pressure of ~ $4\times10^{-6}$ mbar, and the deposition pressure was maintained at ~ $6\times10^{-3}$ mbar. For the reactive sputtering of the TiN electrodes, nitrogen gas was introduced at a flow rate of 2 sccm.

***TEM sample preparation:*** All TEM lamellae were extracted directly from the fully fabricated device electrodes using a Thermo Scientific Helios 5 DualBeam focused ion beam (FIB) system. The lamellae were prepared using a standard FIB lift-out process. To protect the region of interest, an initial electron-beam Pt layer was deposited, followed by ion-beam Pt deposition prior to bulk milling. The initial thinning was performed at 30 kV with progressively lower currents (from 90 pA down to 45pA) Final cleaning and polishing were carried out at lower accelerating voltages (5 kV, followed by 2 kV) and reduced currents (~24 pA) using a combination of high tilt angles ±3° and ±5° for approximately 1 minute on each side of the lamella to ensure minimal beam-induced damage.

**<u>Characterization tools</u>**

***XRD:*** Rigaku SmartLab X-ray diffractometer with high resolution 1.5406 Å Cu-$K_a$ monochromatic radiation is used for crystal structure, phase and orientation analysis.

***Transport measurement***: Ferroelectric tester from Radiant Technologies is used to measure polarization and instantaneous current.

***TEM***: All TEM images were taken with TEM TITAN THEMIS 300 with an acceleration voltage of 300 kV in STEM HAADF mode with a convergence angle of 24.5 mrad and with HAADF collection angle of 48-196 mrad. The TEM is also equipped with Super-X EDS detector with quad segments and EDS spectra were acquired until sufficient counts were obtained.

***XPS:*** Axis Supra$^+$ from Kratos Analytical with monochromatic Al $K_a$ (1486.6 eV) is used to collect XPS spectra. Pass energy for narrow and wide scan are 20 eV and 160 eV respectively. Slot aperture of 300µm × 400 µm and the hybrid mode lens is used.

***ToF-ERDA:*** The samples were analyzed using the ToF-ERDA set up at the Surrey Ion Beam Centre [62]. With the incident ions ($^{127}I$) produced using a HV 860C multi-sample sputter source, using AgI solid target to produce $^{127}I^-$. The negative ions are switched to positive ions and accelerated to 14 MeV $^{127}I^{7+}$, using a HV 2MV Tandetron accelerator, by applying a terminal voltage and stripper gas, $N_2$. Samples were measured at incident and exit angles of 69.5° and a scattering angle of 41°.

**Acknowledgements**

The authors gratefully acknowledge Prof. Anuj Goyal of the Indian Institute of Technology Hyderabad for insightful discussions on defect chemistries and reactions. The authors also acknowledge the Micro and Nano Characterization Facility (MNCF) and the National Nanofabrication Centre (NNfC) at the Centre for Nano Science and Engineering (CeNSE), Advanced Facility for Microscopy and Microanalysis (AFMM), Indian Institute of Science (IISc), Bengaluru, and Indus-I BL-2, RRCAT, Indore, India, for access to fabrication and characterization facilities. The authors are grateful to Sanjoy Kr Mahatha, Abhishek Chaturvedi, and Sinchana R for their technical assistance during XPS measurements. The ToF-ERDA measurements were supported through the Surrey Ion Beam Centre, which is funded by the UK EPSRC (EP/X015491/1). The authors thank Vlad Palitsin for his help in the ion beam analysis laboratory at the Surrey Ion Beam Centre. This work was supported by funding from SERB (DST), New Delhi, Government of India (CRG/2022/003506), as well as support from the DST-COE on piezoMEMS (DST/TDT/AM/2022/084).

**Declaration:** LLMs such as ChatGPT/Gemini was used solely to improve grammar, sentence structure, and language clarity where necessary. These were not used to generate scientific content, interpret data, perform analyses, or produce any results or conclusions.

**Anion Doping Driven Non-Ferroelectric-to-Ferroelectric Phase Transition in Epitaxial $Y:HfO_2$**

Soumyajyoti Mondal [a], Binoy Krishna De [a], Asraful Haque [a], Shubham Kumar Parate [a], Arup Basak [a], Naushad Ahemad [a], Pramod Kumar Yadav [a], Kaushal Tiwari [b], Bhagwati Prasad [b], Matthew K. Sharpe [c], Catia Costa [c], Satheesh Krishnamurthy [c], Pavan Nukala [a,b,*]

[a] *Centre for Nano Science and Engineering, Indian Institute of Science, Bengaluru, 560012, India.*

[b] *Department of Materials Engineering, Indian Institute of Science, Bengaluru, 560012, India.*

[c] *Surrey Ion Beam Centre, University of Surrey, Guildford, Surrey, GU2 7XH, United Kingdom.*

* *Corresponding author* Pavan Nukala, e-mail: pnukala@iisc.ac.in

a)

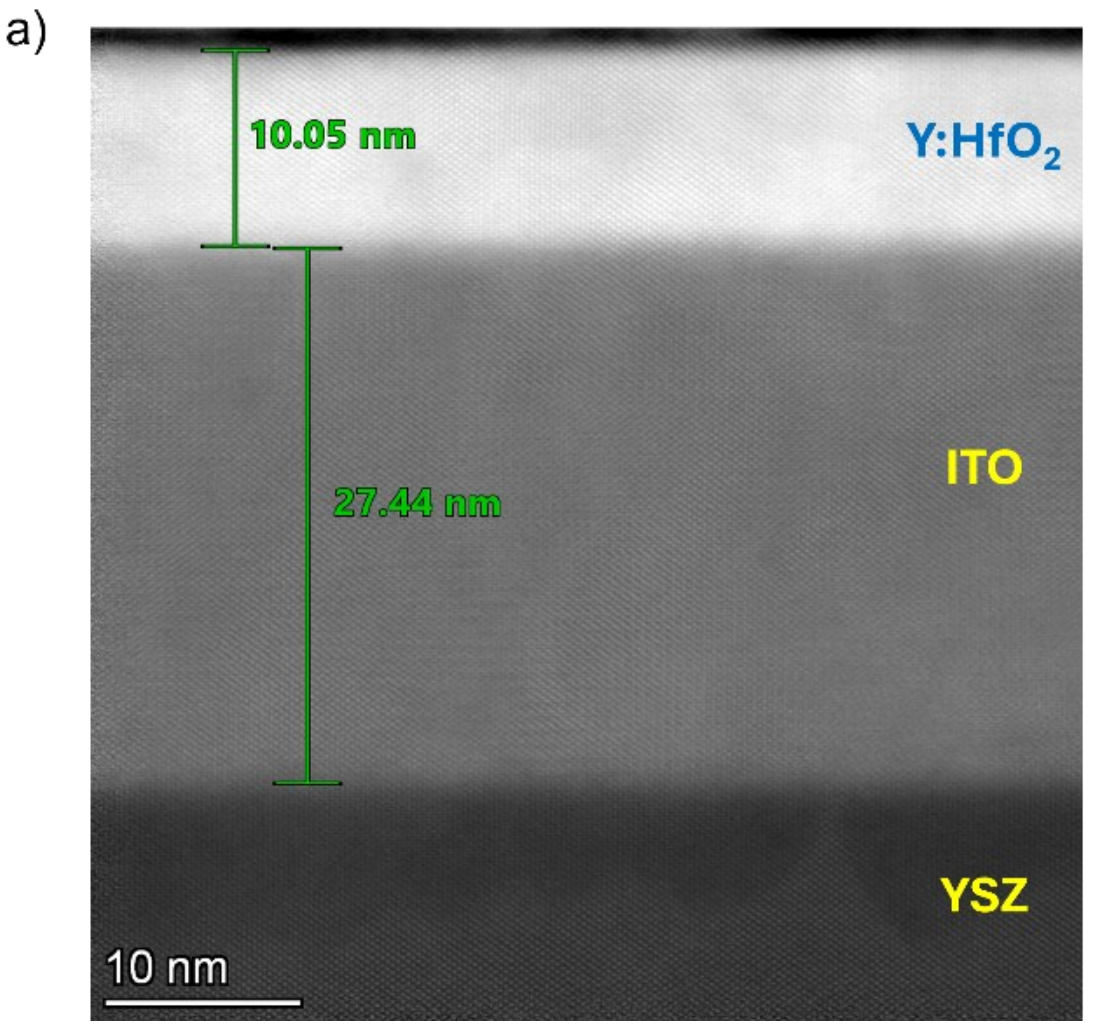

b)

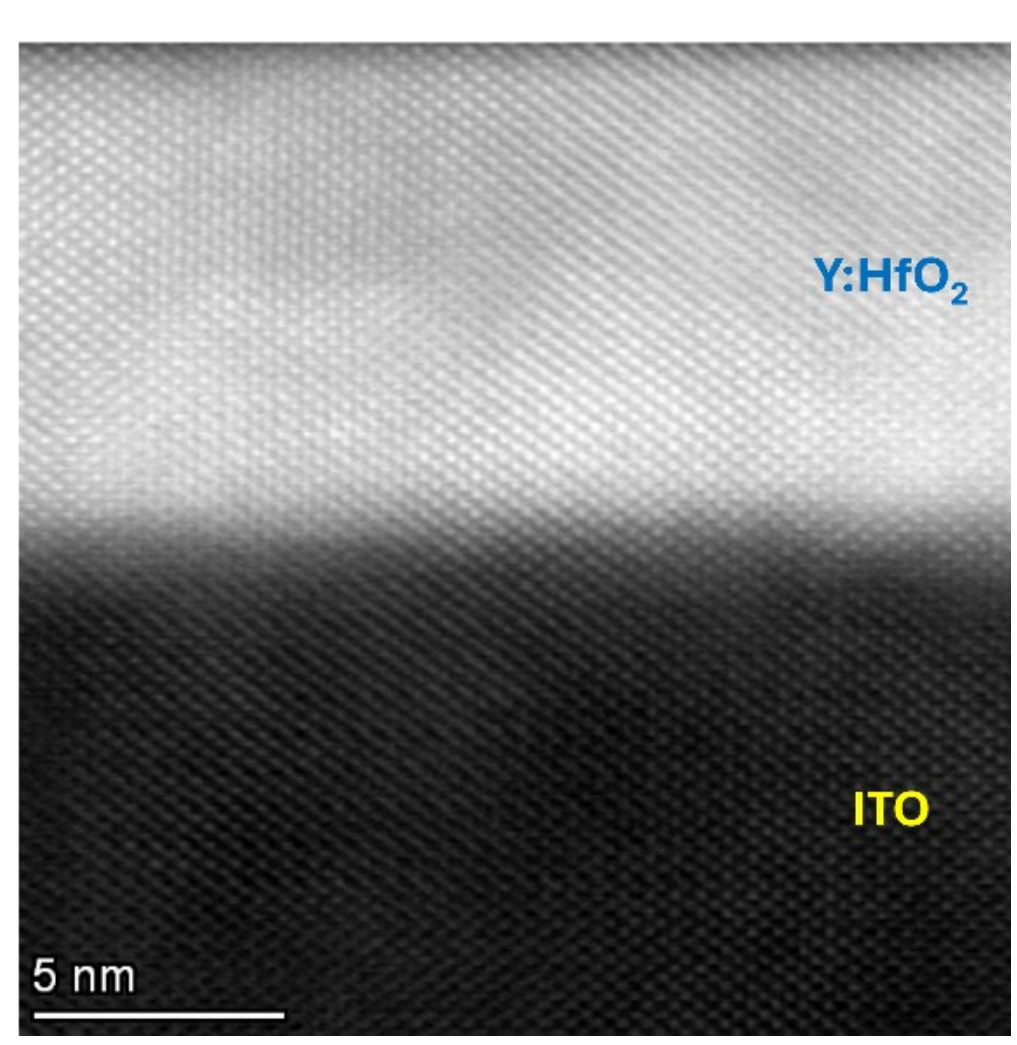

**Figure S1:** **HAADF-STEM analysis of interfaces** a) Cross-sectional HAADF-STEM image of the as deposited ferroelectric $Y:HfO_2$ film. b) Zoomed-up section of interface between $Y:HfO_2$ and ITO.

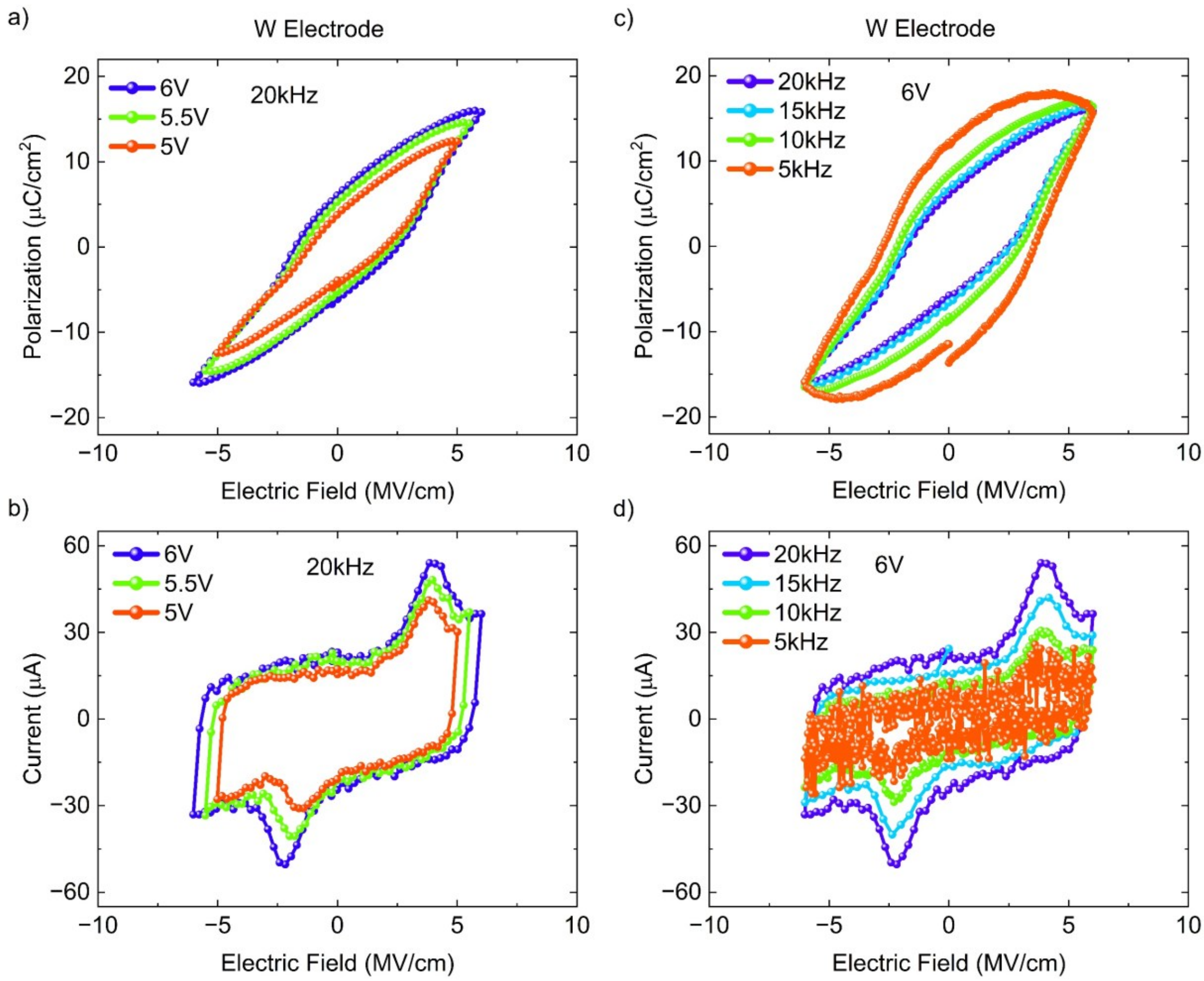

**Figure S2: Electrical characterization of the as deposited ferroelectric film of thickness ~10nm interfaced with W top electrodes**. a) Polarization and b) Instantaneous current vs electric field plot for various applied voltage at 20 kHz frequency; c) Polarization and b) Instantaneous current vs electric field plot at 6 V applied voltage measured at various frequencies.

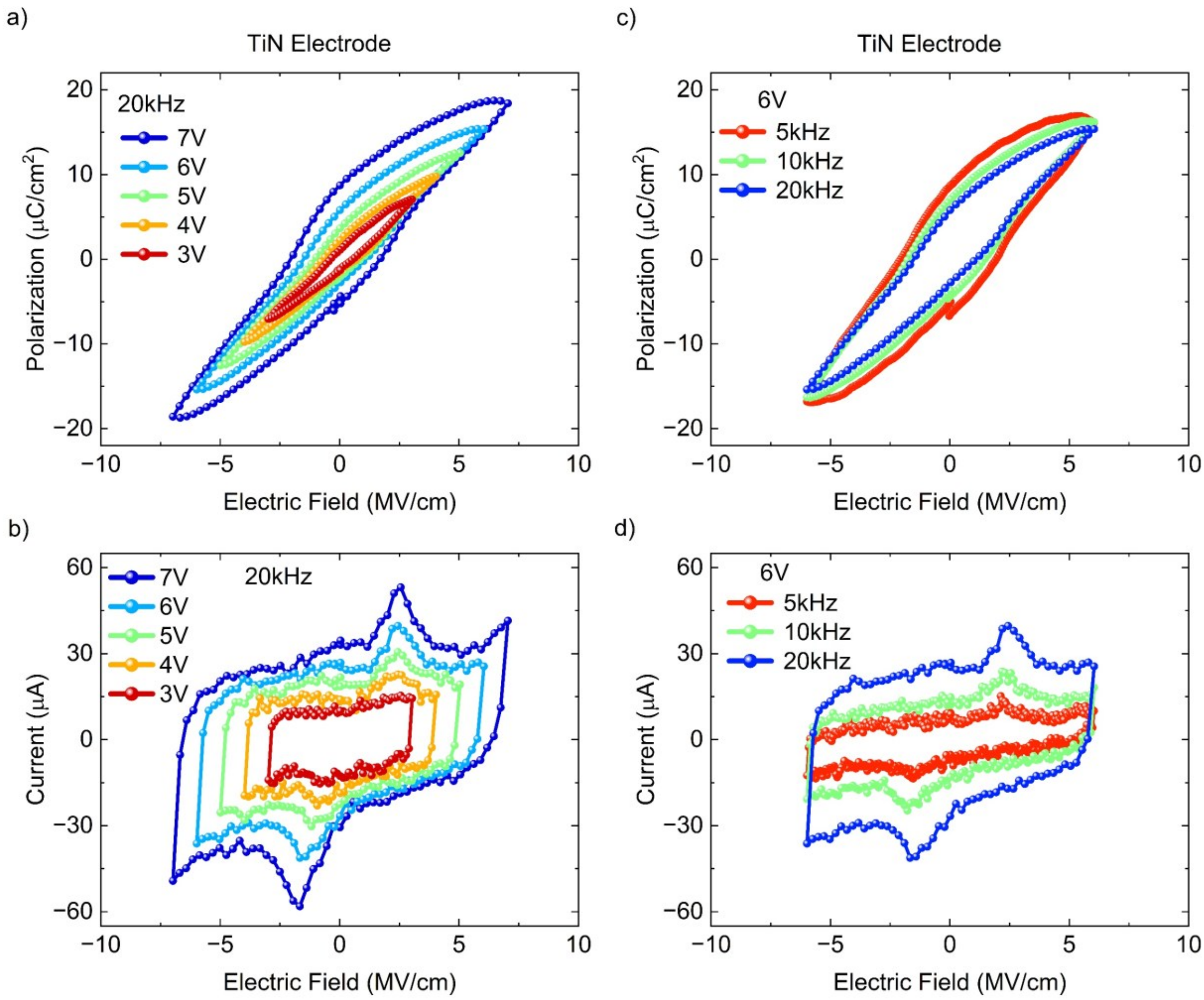

**Figure S3: Electrical characterization of the as deposited ferroelectric film of thickness ~10nm interfaced with TiN top electrodes.** a) Polarization and b) Instantaneous current vs electric field plot for various applied voltage at 20 kHz frequency; c) Polarization and d) Instantaneous current vs electric field plot at 6 V applied voltage measured at various frequencies.

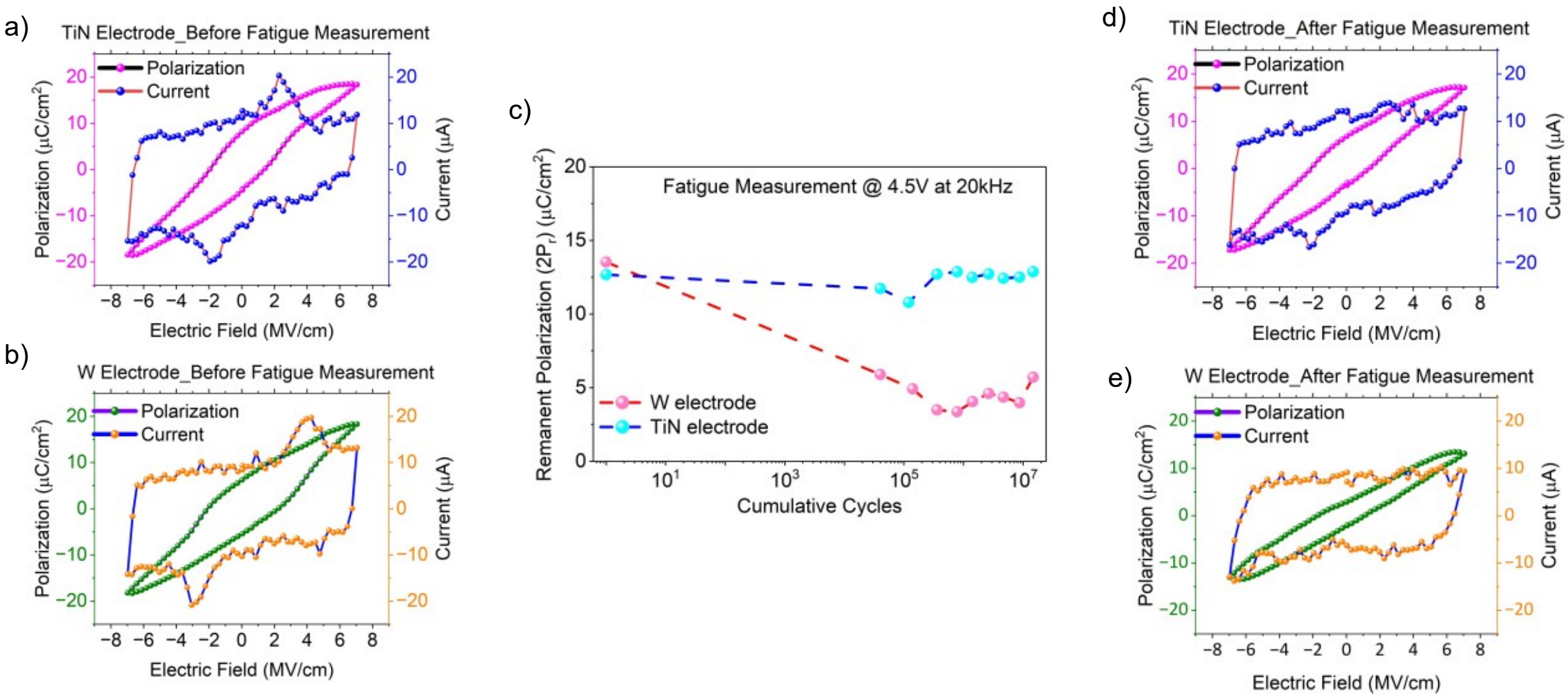

**Figure S4: Endurance test:** Polarization and current vs electric field plot for as deposited ferroelectric film interfaced with a) TiN and b) W top electrodes (30μm diameter) before

fatigue measurements; c) Remanent polarization ($2P_r$) as a function of cumulative number of cycles during fatigue measurement of both the devices, performed at 4.5 V and 20 kHz; Polarization and current vs electric field plot for the film interfaced with d) TiN and e) W top electrode after cycling.

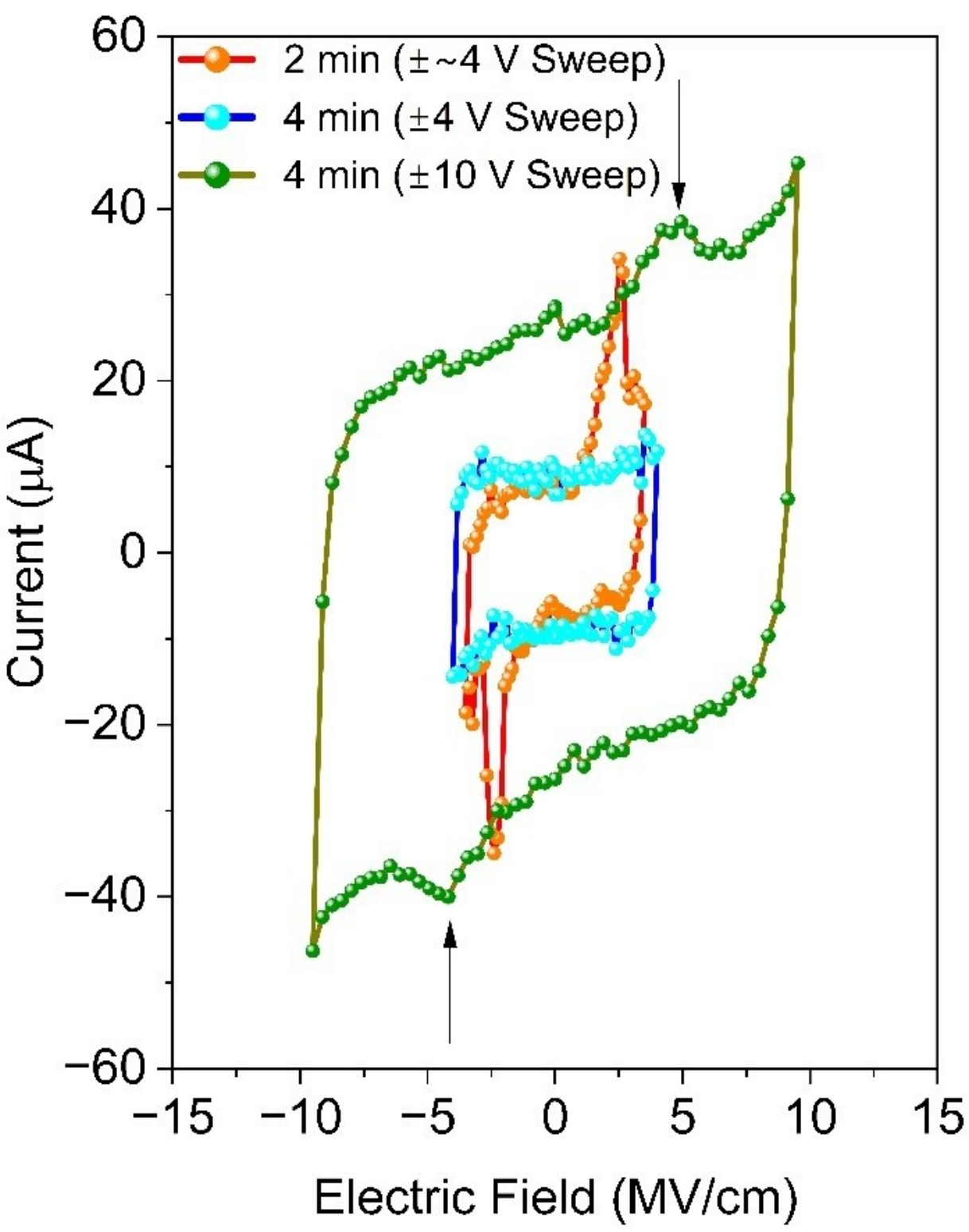

**Figure S5: Current-electric field characteristics of RTA-treated samples.** The plot displays the field-dependent current response for varied RTA exposure times (2 min and 4 min). The black arrows highlight the ferroelectric current switching peaks observed in the 4-minute RTA sample.

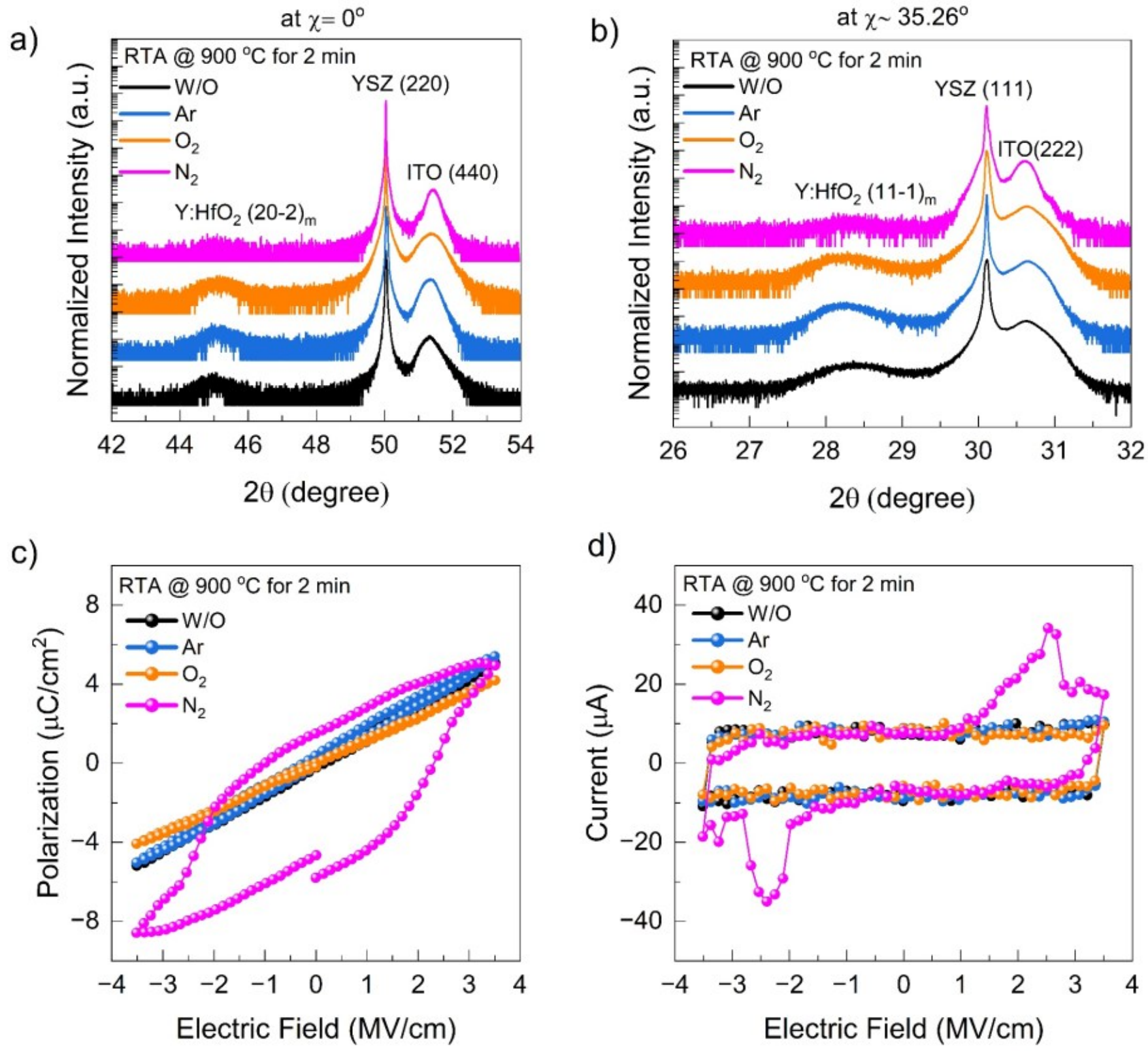

**Figure S6:** Effect of rapid thermal annealing at various ambience (Ar, $O_2$ and $N_2$) at 900 ºC for 2 min on as grown non-ferroelectric sample. a) XRD θ-2θ scan at χ= 0º, b) XRD θ-2θ scan at χ= 35.26º, c) Polarization and d) Instantaneous current as function of electric field measured at 20 kHz frequency for the devices interfaced with W top electrodes.

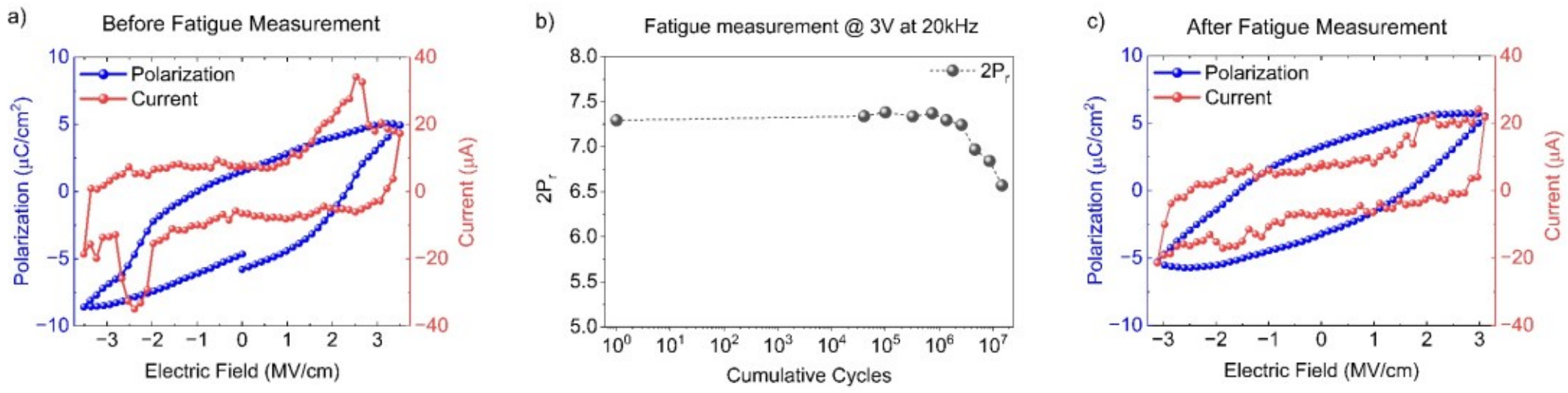

**Figure S7:** Endurance test on the RTA-treated $Y:HfO_2$ film: **(a)** Polarization and current versus electric field (P-E and I-E) loops before fatigue measurements; **(b)** Remanent polarization ($2P_r$) as a function of the cumulative number of cycles during fatigue testing, performed at 3 V and 20 kHz; **(c)** P-E and I-E loops for the film after completing the fatigue cycling.

## Text S1:

Our films exhibit lower measured $P_r$ values compared to the theoretically predicted spontaneous polarization values reported for orthorhombic hafnia systems [1,2] . As discussed by Tan et al.[3] for epitaxial hafnia-based ferroelectric films, the experimentally measured ferroelectric response strongly depends on the projection of the polarization vector along the measurement direction. Therefore, in epitaxial hafnia systems, the measured out-of-plane $P_r$ may underestimate the actual orthorhombic ferroelectric phase fraction due to orientation-dependent projection effects. Although the epitaxial relationship in the present YSZ (110)-based films differs from the STO-based systems discussed in the cited works, the general projection argument remains relevant for interpreting the measured polarization response. In addition, the residual monoclinic phase fraction present in the films can further reduce the measured polarization values in our system. While certain epitaxial systems (such as those grown on STO) exhibit larger polarization values, the literature also demonstrates that the measured $P_r$ is highly sensitive to the substrate choice, epitaxial orientation (projection effects), and the coexistence of residual monoclinic phases. To better contextualize our findings, we have incorporated the highly relevant literatures [1,3–9] in **Table 1**.

**Table 1:**

| S. No. | Group / Reference | Nature of Film | Nature of Study | Substrate / Geometry | Dominant Orientation | Reported $P_r$ ($\mu C\ cm^{-2}$) | Relevance to Present Work |
|---|---|---|---|---|---|---|---|
| 1 | Huan et al. / Phys. Rev. B 90, 064111 [1] | — | Theoretical | Orthorhombic $HfO_2$ | — | ~52–56 ($P_s$) | Predicts the spontaneous polarization ($P_s$), establishing the theoretical upper limit which is fundamentally larger than experimentally measured remanent values. |
| 2 | Mimura et al. / Appl. Phys. Lett. 113, 102901 [4] | Epitaxial | Experimental | $Y{:}HfO_2$ / ITO(111)// YSZ(111) | Predominantly (111)-oriented orthorhombic phase | ~12 | Shows stable ferroelectricity on ITO/YSZ, serving as the standard baseline for evaluating our (110)-oriented films on similar substrates. |
| 3 | Mimura et al. / Jpn. J. Appl. Phys. | Epitaxial | Experimental | $Y{:}HfO_2$ / ITO(111)// YSZ(111) | (111)-oriented orthorh | ~12–15 | Highlights that measured $P_r$ values are strongly |
| | 58, SBBB09 [5] | | | | ombic/tetragonal | | limited by residual monoclinic phases, supporting our observations of phase competition. |
| 4 | Mimura et al. / Appl. Phys. Lett. 115, 032901 [6] | Polycrystalline | Experimental | $Y:HfO_2$ / $Pt/TiO_x/SiO_2/Si$ | Orthorhombic/tetragonal dominant | ~14–17 | Demonstrates stable ferroelectricity in $Y:HfO_2$, contrasting typical polycrystalline defect dynamics with our single-crystalline epitaxial approach. |
| 5 | Yun et al. / Nature Mater. 21, 903–909 [7] | Epitaxial | Experimental | $Y:HfO_2$ / LSMO// STO | (111)-oriented epitaxial film | ~32 | Demonstrates high polarization via cationic defects, contrasting with our novel RTA-induced nitrogen (anionic) doping mechanism. |
| 6 | Mimura et al. / Phys. Status Solidi A 220, 2300100 [8] | Epitaxial | Experimental | x%$YO_{1.5}$–(100−x%)(Hf, Zr)$O_2$ / (111)YSZ | (111)o/(101)t-oriented epitaxial films | ~12–19 | Shows ferroelectric response despite mixed phases, contextualizing our measured polarization in the presence of suppressed monoclinic fractions. |
| 7 | Tan et al. / J. Mater. Chem. C 11, 7219–7226 | Epitaxial | Experimental | HZO / STO | Tilted (111)-oriented | Orientation dependent | Crucial for interpreting our data: proves that |

| | [3] | | | | epitaxial domains | | measured $P_r$ depends heavily on the polarization vector projection, justifying the values seen in our (110) geometry. |
|---|---|---|---|---|---|---|---|
| 8 | Farahani et al. / ACS Appl. Electron. Mater. 7, 6628–6634 [9] | Epitaxial | Experimental | $Y:HfO_2$ / STO(001), STO(110) | o(111)-dominant films with monoclinic phase coexistence | ~8–33 | Reports that even in highly optimized Y-doped epitaxial films, a paraelectric monoclinic fraction inherently coexists. This thermodynamically validates our observations, showing that while N-doping successfully drives our monoclinic as-grown film into a dominant orthorhombic phase, a minor residual monoclinic fraction is typical for this material system. |
| 9 | Present work | Epitaxial | Experimental | $Y:HfO_2$ / ITO//YSZ(110) | (110)-oriented | ~3.67 (N-doped) ~6.0 (As-grown PLD) | Proves that anionic N-doping successfully induces a ferroelectric transition. Our measured $P_r$ is consistent with |

| | | | | | | | the expected geometric projection for a (110)-oriented system, while the presence of a minor residual monoclinic phase also contributes to the reduced polarization value. |
|---|---|---|---|---|---|---|---|

**Text S2: XPS Peak Deconvolution and Fitting Procedure**

The intensities (areas under the curve) of the non-overlapping Hf 4f (~18 eV), O 1s (529.4 eV), and Y 3d (~158 eV) peaks were extracted by fitting the XPS spectra using XPSPEAK41 software. These peaks were used to determine the atomic fractions of Hf, O, and Y, respectively. The obtained areas and atomic fractions are summarized in **Table S2.**

The nitrogen concentration is determined using the N 1s peak. However, the Y 3s (394.8 eV) and N 1s (397–398 eV) peaks are close in binding energy. To properly deconvolve these peaks, we first calculated the expected intensity of the Y 3s peak based on the atomic fraction obtained from the well-resolved Y 3d peak (**Table S2**). This calculation was performed using the following relationship [10]:

$$\frac{A_{Y\ 3s}}{RSF_{Y\ 3s} \times T_{\ 3s}} = \frac{A_{Y\ 3d}}{RSF_{Y\ 3d} \times T_{\ 3d}}$$

where I is the peak intensity (area), RSF is the relative sensitivity factor (referenced to C 1s), and T is the transmission factor of the instrument.

The spectra around 400 eV were then fitted with two peaks corresponding to Y 3s and N 1s, following a Shirley background subtraction. During this fitting process, the peak area of Y 3s was fixed to the calculated value to isolate the true N 1s peak area. Fitted peak parameters are listed in **Table S3**. The resulting N 1s peak areas and N/(Y+Hf) ratio for samples treated with different Rapid Thermal Annealing (RTA) times are provided in **Table S2**.

**Table S2:** Atomic fraction calculation

RSF - 2.93 (O), 1.8(N), 7.52 (Hf), 5.98 (Y 3d) and 1.98(Y 3s); instrument transmission factor T - 2.853209 (O), 2.781117(N), 1.988309 (Hf) and 2.378365 (Y 3d), 2.779626 (Y 3s) (kinetic energy dependent response of the detector).

| Sample | $I_{Hf 4f}$ | $I_{O1s}$ | $I_{Y3d}$ | $I_{Y3s}$ | $I_{N1s}$ | Y/(Y+Hf) % | O/(Hf+Y) ratio | N/(Y+Hf) ratio |
|---|---|---|---|---|---|---|---|---|
| W/O RTA | 47685 | 45893 | 4208 | 1393 | 0 | 8(0.3) | 1.57(0.06) | 0 |
| 10 s RTA | 66172 | 68424 | 6055 | 2004 | 1570 | 9(0.2) | 1.68(0.04) | 0.05(0.03) |
| 1 min RTA | 52434 | 50677 | 4490 | 1486 | 1804 | 8(0.2) | 1.58(0.055) | 0.07(0.04) |
| 2min RTA | 45822 | 46979 | 4349 | 1685 | 1745 | 9.(0.3) | 1.66(0.05) | 0.08(0.05) |
| 4 min RTA | 49576 | 45908 | 4880 | 1616 | 1337 | 9(0.3) | 1.50(0.04) | 0.06(0.05) |

**Table S3:** Parameters of Y 3s and N1s peaks

| Sample | Y 3s | | | N 1s | | |
|---|---|---|---|---|---|---|
| | BE (eV) | $I_{Y3s}$ | FWHM (eV) | BE (eV) | $I_{N1s}$ | FWHM (eV) |
| w/o RTA | 395.1 | 1372 | 3.0 | | | |
| 10 s RTA | 394.2 | 2003 | 2.96 | 396.9 | 1570 | 3.9 |
| 1 min RTA | 394.8 | 1486 | 2.95 | 397.6 | 1804 | 4.2 |
| 2 min RTA | 394.9 | 1685 | 2.96 | 398.3 | 1745 | 4.4 |
| 4 min RTA | 394.8 | 1616 | 2.95 | 397.8 | 1338 | 3.8 |

## 2. Detection Limit and Confidence Level Calculations

The reliability of the N 1s feature was rigorously tested by evaluating the statistical level of confidence [11] and calculating the detection limit of nitrogen under our experimental conditions. In XPS, a spectral feature is generally considered a genuine peak if the confidence level exceeds 95%.

The parameters used to verify the confidence level of the N 1s peaks are listed in **Table S4**. The measured peak areas are greater than the minimum area required for a 95% confidence level, verifying the reliability of the N 1s spectral feature at ~397 eV. Furthermore, the detection limit of N was calculated using the methodology reported by Shard [11]. The observed nitrogen percentages in our films are strictly above this calculated detection limit (**Table S4**).

**Table S4:** Limit of detection (LOD) and level of confidence test.

| | N | W | $\sigma_\beta$ | Area for 95% confidence $I_{LOD}(N\ 1s) = 2\sigma_\beta\sqrt{N(1+\frac{N}{2W})}$ | Obtained area form fitting | LOD $X_{LOD}(N\ 1s) = \frac{\sigma_\beta}{\sigma_\beta + (\varepsilon^{0.5}.RSF_N.T_N/k)(I_{Hf}/RSF_{Hf}.T_H}$ |
|---|---|---|---|---|---|---|
| 10 s RTA | 112 | 29 | 34 | 1232 | 1570 | 4.1% |
| 1 min RTA | 130 | 36 | 38 | 1451 | 1804 | 5.8% |
| 2min RTA | 134 | 33 | 33 | 1344 | 1745 | 5.7% |
| 4 min RTA | 129 | 33 | 27 | 1073 | 1337 | 4.4% |

N- Number of points used to determine the peak area

W – Number of points used to determine $\sigma_\beta$.

$\sigma_\beta$ – Standard deviation of background

ε - Step size

k – Detection limit scaling factor, used here = 9, ref. [11].

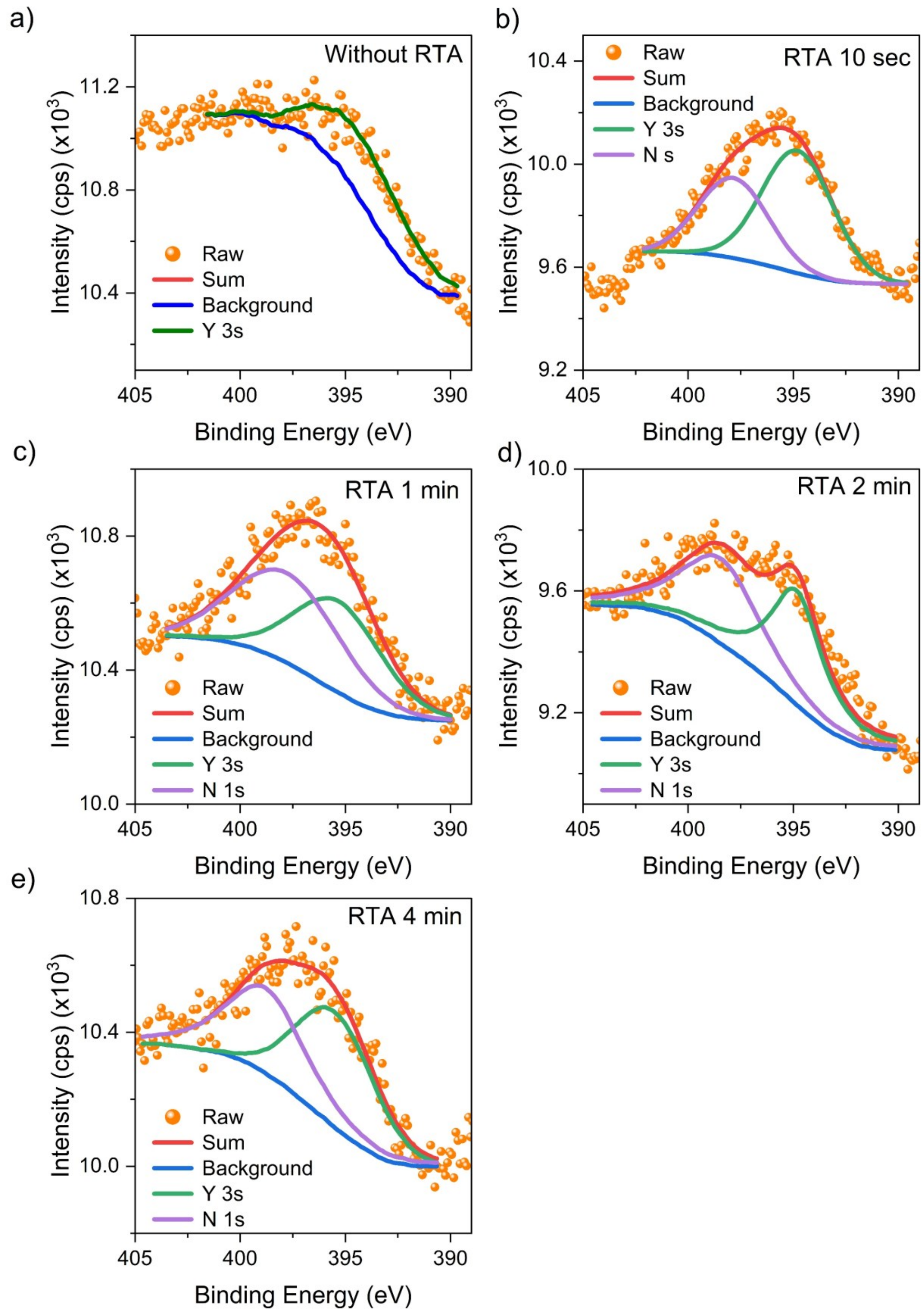

**Figure S8**: **XPS spectra from as deposited non-ferroelectric Y:HfO$_2$ film undergone various durations of RTA exposure and, Y 3s and N1s peak fitting.** Intensity vs binding

energy plot for a) Without RTA sample, b) 10 second, c)1 minute d) 2 minutes e) 4 minutes of RTA exposure.

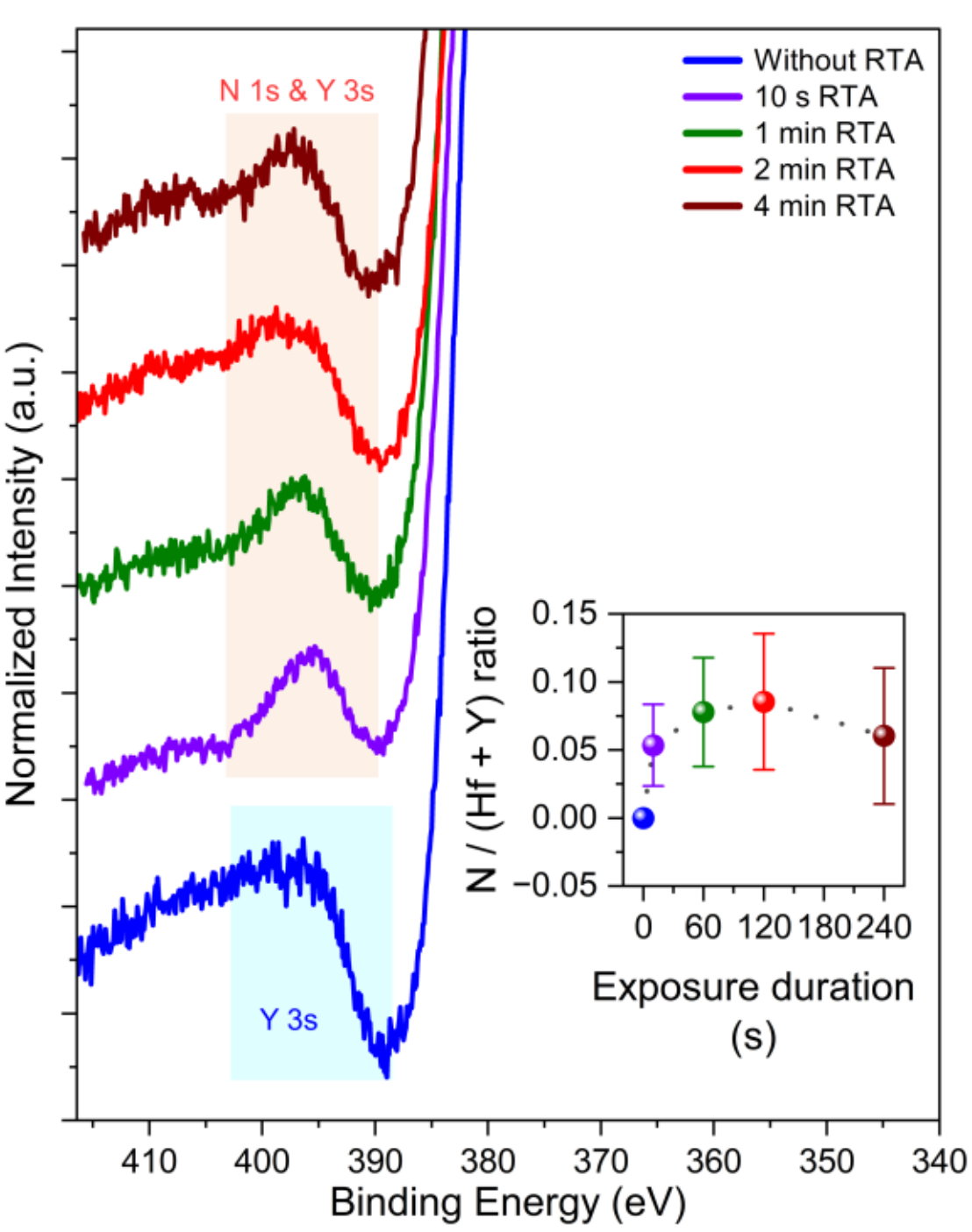

**Figure S9:** XPS spectra from Y:$HfO_2$ samples RTA treated for various durations of exposure. N to total cation ratio calculated from XPS as a function of RTA exposure duration (inset).

## Text S3:

While it is clear that N content (N/Hf+Y) increases from non-RTA to RTA processed samples, it is unclear whether it decreases from 2 min to 4 min samples. The estimated N content fall within the error bars. Given that we keep supplying $N_2$, it is unlikely that N concentration reduces from 2 min to 4 min RTA processed samples. As a result, we propose the following mechanism: N doping initially quenches neutral oxygen vacancies creating charged oxygen vacancies, which transform the monoclinic phase into polar orthorhombic phase. The presence of monoclinic phase also increases the coercive field for switching in 4 min RTA samples (**Figure S5**). It is likely that beyond a certain concentration of N doping, N will start to replace oxygen sites (and not neutral vacancies), and can create neutral vacancies again, that will restabilize the monoclinic phase.

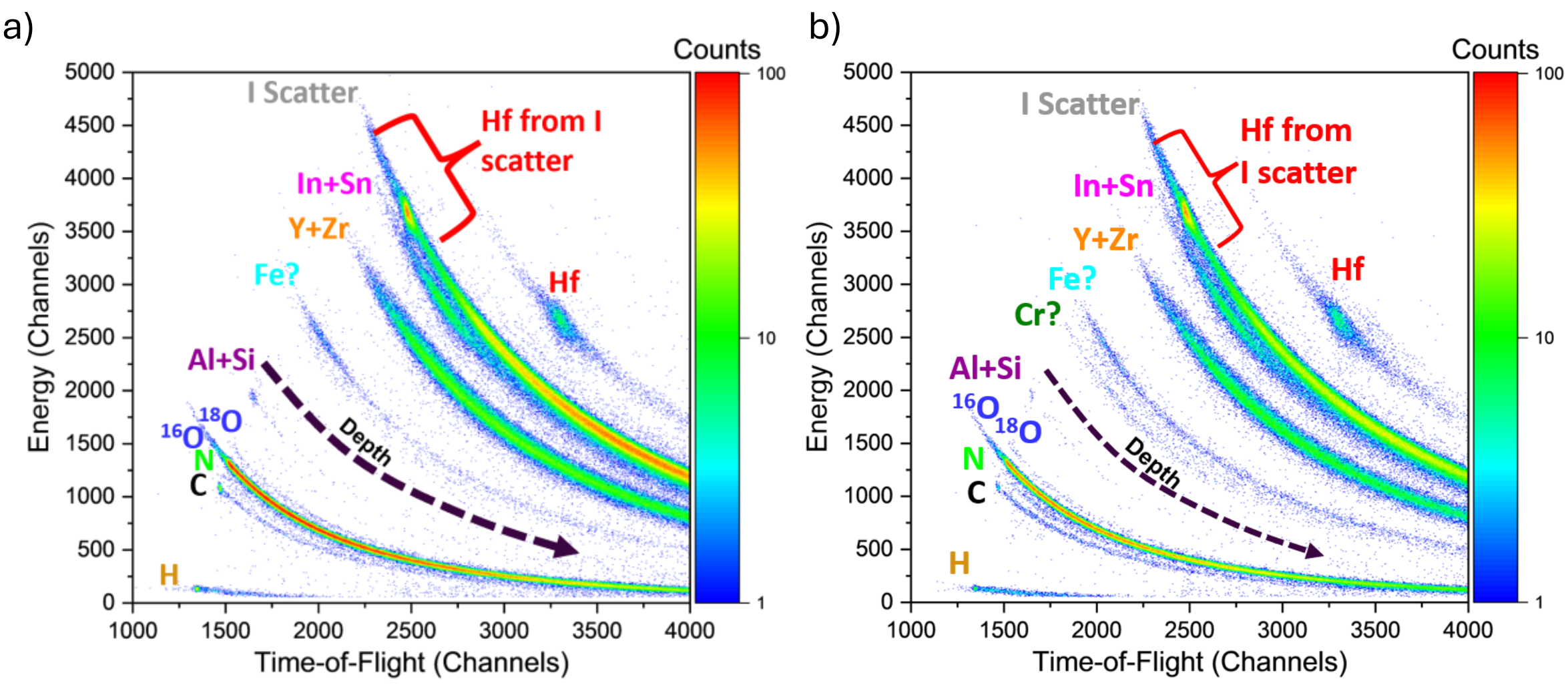

**Figure S10:** ToF-E histograms and depth profiles from the ToF-ERDA measurements. Panel a) shows the ToF-E histogram for the without RTA sample and b) shows the ToF-E histogram for the 2 min RTA sample. The concentrations in the depth profiles are shown on a log scale.

**Table S5:** Layer compositions and thicknesses from the ToF-ERDA data fitted with MCERD simulations for the major elements in both the without RTA and 2 min RTA sample.

Without RTA:

| **Density / g/cm³** | **Thickness / nm** | **O /at.%** | **Y /at.%** | **Y+Zr /at.%** | **In+Sn /at.%** | **Hf /at.%** |
|---|---|---|---|---|---|---|
| 9.7 | 10 | 66.23 | 2.65 | - | - | 31.13 |
| 7.14 | 34 | 64.81 | - | - | 35.19 | - |
| 6 | Substrate | 69.23 | - | 30.77 | - | - |

2 min RTA:

| **Density / g/cm³** | **Thickness / nm** | **O /at.%** | **Y /at.%** | **Y+Zr /at.%** | **In+Sn /at.%** | **Hf /at.%** |
|---|---|---|---|---|---|---|
| 9.7 | 10 | 70.39 | 2.69 | - | - | 26.92 |
| 7.14 | 34 | 66.67 | - | - | 33.33 | - |
| 6 | Substrate | 68.09 | - | 31.91 | - | - |

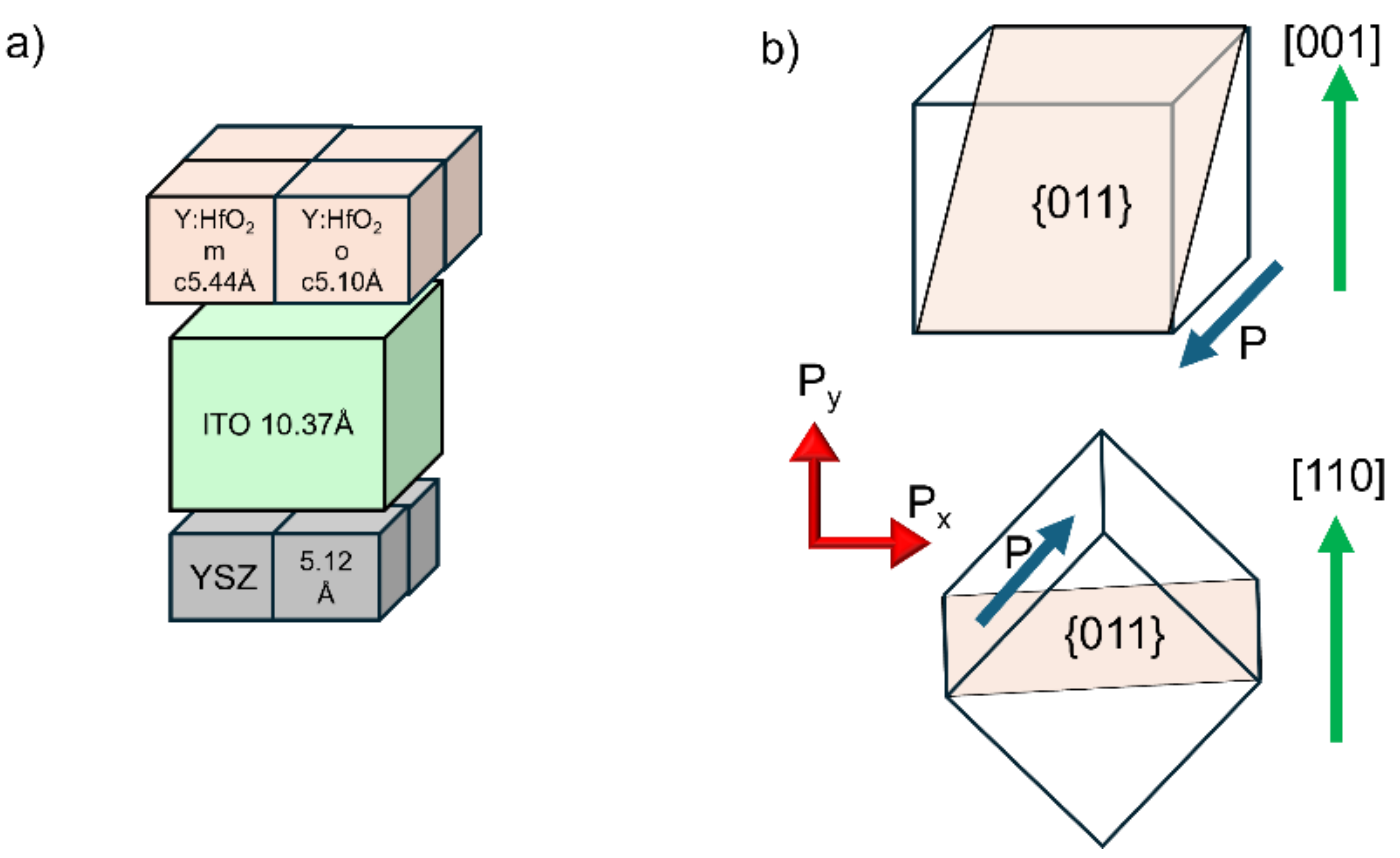

**Figure S11:** **Epitaxial relationship between YSZ, ITO and LHO.** a) Schematic of one unit cell of each layer in out of-plane direction. One unit cell of ITO matches with four-unit cells of YSZ and Y:HfO$_2$, and maintains the in-plane registry. b) Schematic of the unit cell of Y:HfO$_2$ as [001] and [110] as out of-plane directions.

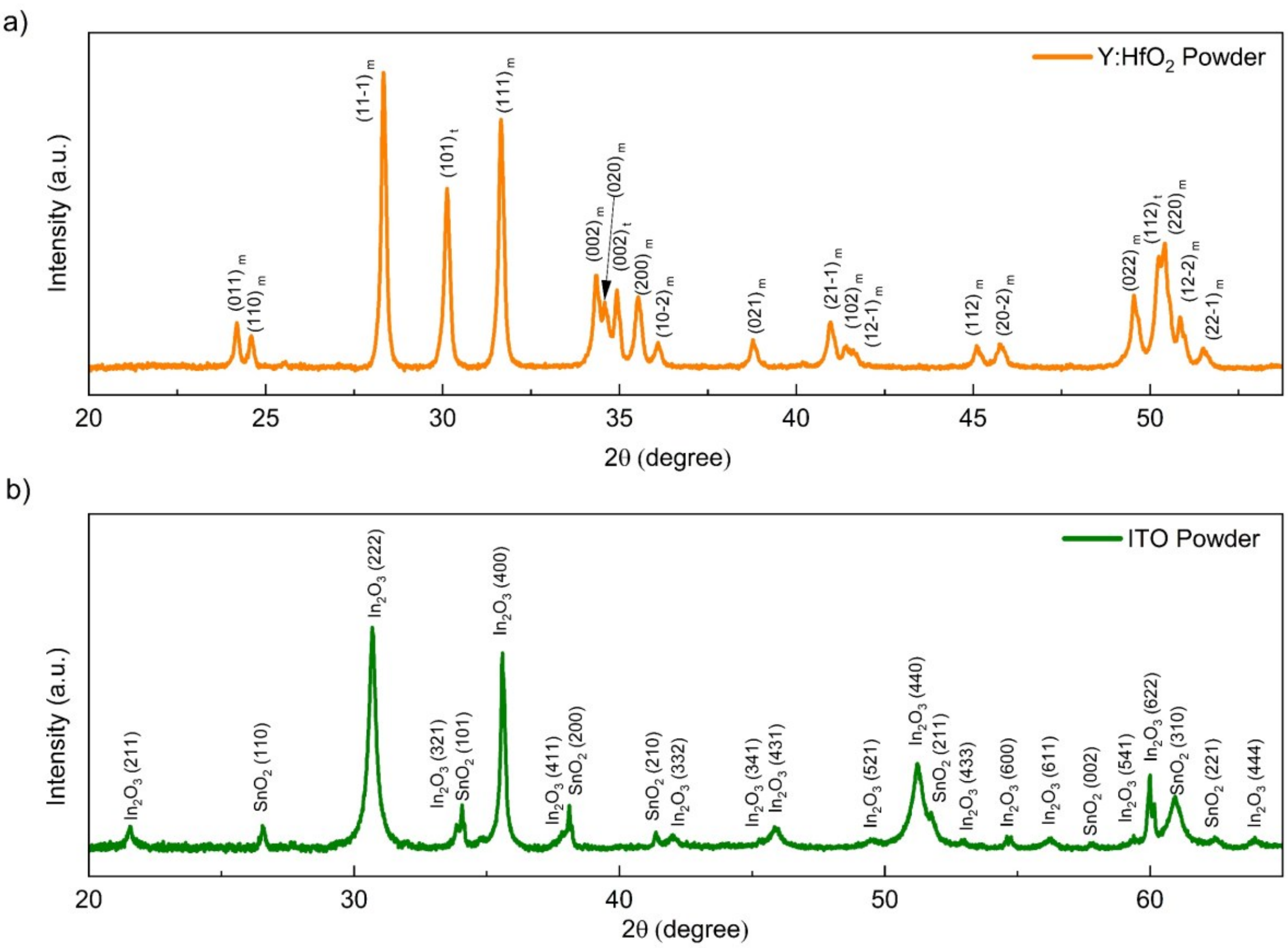

**Figure S12:** XRD θ-2θ scan of the powder scratched from the PLD targets a) the solid-state synthesized ~7% Y: $HfO_2$ target and b) the commercial ITO target

.

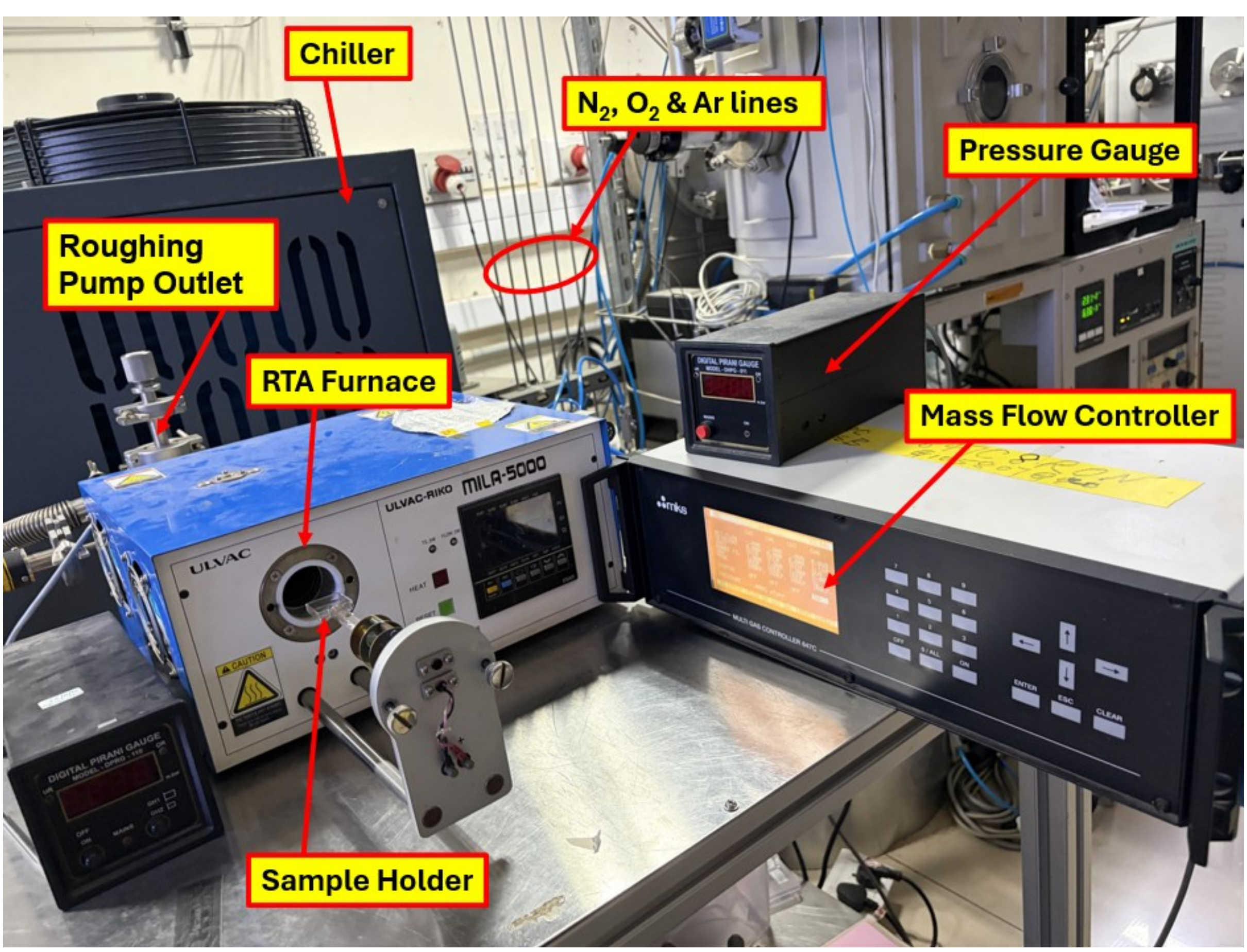

**Figure S13:** Custom-built rapid thermal annealing tool.

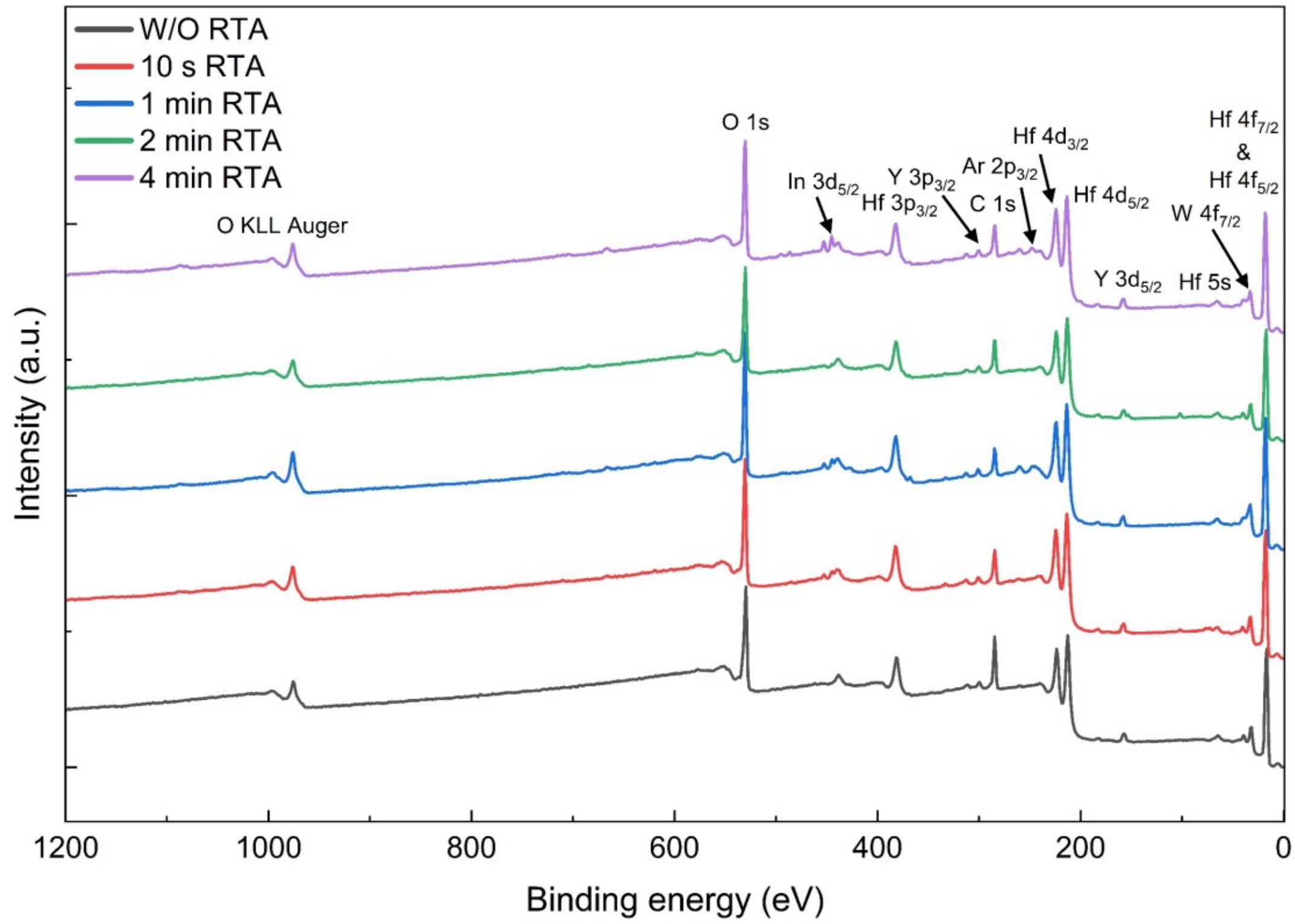

**Figure S14:** XPS survey spectra from the RTA treated samples for various duration of exposure in $N_2$ ambience at 900 °C.

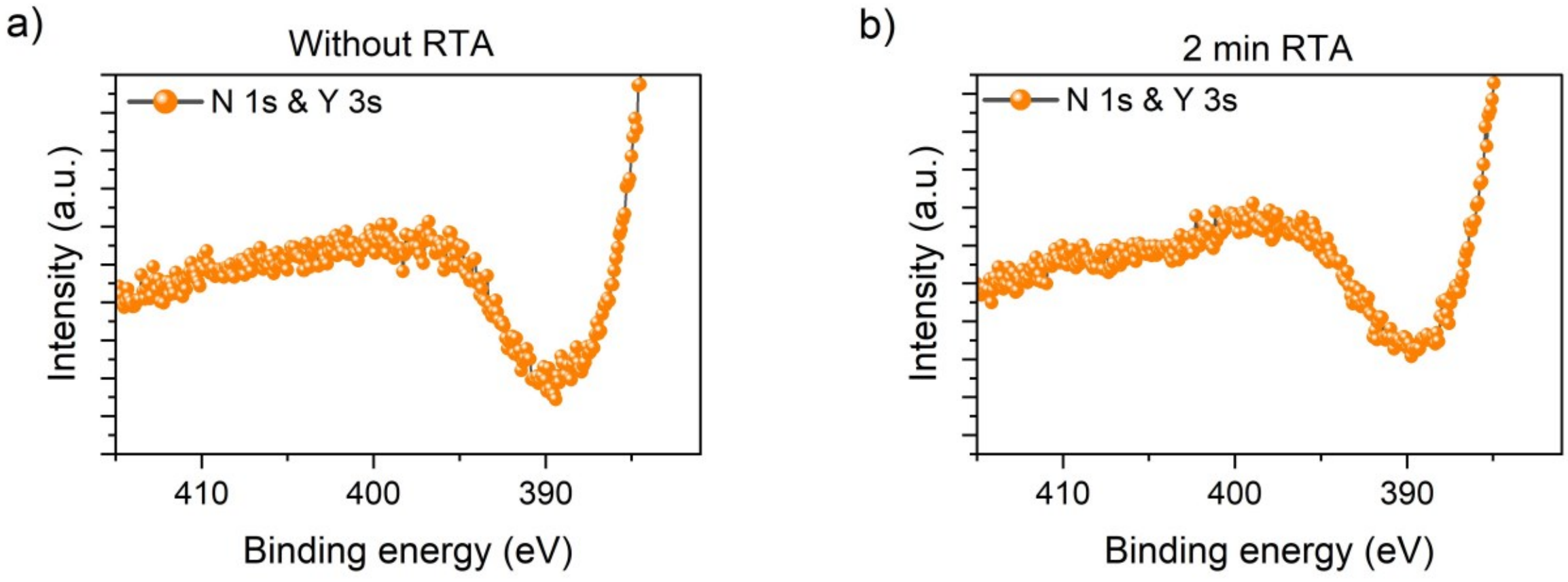

**Figure S15:** High-resolution XPS spectra of the overlapping N 1s and Y 3s core-level regions for the epitaxial Y:$HfO_2$ thin films. Panel (a) shows the as-grown sample without RTA, and panel (b) shows the sample after a 2-minute RTA treatment. The binding energies are charge-corrected using the C 1s peak at 284.6 eV.